\documentclass{aa}
\usepackage{natbib}
\usepackage{amsmath,amssymb,epsfig}
\usepackage{graphicx}
\makeatletter
\def\mycaption#1#2{{\if#1a\global\advance\c@figure1\fi\global\advance\c@figure-1\let\savethefigure\thefigure\def\thefigure{\savethefigure#1}\caption{#2}}\protected@edef \@currentlabel {\csname p@figure\endcsname \csname thefigure\endcsname#1}}
\makeatother

\newcommand{\rt}{({\bf x},t)}
\newcommand{\mbm}[1]{\mbox{\boldmath$#1$}}
\begin{document}
\title{A Free-Electron Laser in the Pulsar Magnetosphere}
\titlerunning{A FEL in the Pulsar Magnetosphere}
\author{P.K. Fung\inst{1} 
        \and J. Kuijpers\inst{2}}
\authorrunning{P.K. Fung \and J. Kuijpers}
\offprints{P.K. Fung, \email{fung@astro.uu.nl}}
\date{May 2, 2004}
\institute{Astronomical Institute, Utrecht University, P.O. Box 80000, 3508 TA Utrecht, The Netherlands \and Department of Astrophysics, University of Nijmegen, P.O. Box 9010, 6500 GL Nijmegen, The Netherlands }
\abstract{
We have studied systematically the free-electron laser in the context of high brightness
pulsar radio emission. In this paper, we have numerically examined the case
where a transverse electromagnetic wave is distorting the motion of a
relativistic electron beam while travelling over one stellar radius ($\approx
10\:\mbox{km}$). For different sets of parameters, coherent emission is generated by bunches of beam
electrons in the radio domain, with bandwidths of 3 GHz. Pulse power often
reached $10^{13}\:\mbox{W}$, which corresponds with brightness
temperature of $10^{30}\:\mbox{K}$. The duration of these pulses is of
the order of nanoseconds. In the context of pulsar radio emission,
our results indicate that the laser can produce elementary bursts of radiation which
build up the observed microstructures of a few tens of microseconds duration.  
The process is sensitive mostly to the beam particles energy, number
density and the background magnetic field, but much less so to the
transverse wave parameters. We demonstrate 
that the operation of a free-electron laser with a transverse electromagnetic wiggler in the pulsar magnetosphere occurs preferably at moderate 
Lorentz factors $\gamma \geq 100$, high beam density $n \gtrsim 0.1\;n_{\textrm{GJ}}(r_\ast)$ where $n_{\textrm{GJ}}(r_\ast)$ is the Goldreich-Julian 
density at a stellar radius $r_\ast$, and finally, at large altitude
where the background
magnetic field is low $B_0 \leq 10^{-2}\;\textrm{T}$. 
\keywords{radio pulsars; coherent emission; free-electron laser; inverse Compton scattering}
}
\maketitle

\section{Introduction}
Although radio pulsars have been discovered now for more than three decades, the generation of the radio 
signal which reveals their existence, is still a puzzle. It is
believed that a pulsar is a rotating neutron star with a high magnetic
field ($B \sim 10^5 - 10^9\:\mbox{T}$). The radio pulses span a frequency
range from $\sim 100\:\mbox{MHz}$ to $\sim 30\:\textrm{GHz}$ and have the
remarkable property that, in a given pulsar, they vary dramatically
from pulse to pulse, while the {\it average} pulse profile is
extremely stable and unique for that pulsar. On one hand, the stability
and uniqueness of the averaged pulse shape for every pulsar imply that the radio pulse is generated well
inside the light cylinder (defined as the cylinder with radius $R_{\textrm{lc}}
= c/\Omega_\ast$, where $\Omega_\ast$ is the rotation frequency of the
pulsar and $c$ is the speed of light). On the other hand, the
variability of successive pulses suggests that the radio emission
process is strongly fluctuating and/or that the acceleration of
particles responsible for the emission is highly non-steady. The
observations indicate that the radiation is emitted more or less
tangential to the open field lines from the magnetic poles and that it
is polarised preferentially in the local plane of the magnetic
field lines and perpendicular to it. Current belief is that radio
emission is related to the development of a cascade of pairs of
electrons and positrons on the open field lines anchored at the magnetic poles. \\ 
An important property of the pulses is their high brightness
temperature $T_{\textrm{b}}$, given by $T_{\textrm{b}} \sim 10^{34} \: \mbox{K} \: F_{\textrm{Jy}} \: d_{\textrm{kpc}}^2\: r_{\textrm{m}}^{-2} \:
\nu_{\textrm{GHz}}^{-2}$, where $F_{\textrm{Jy}}$ is the measured flux in Jansky at the
frequency $\nu_{\textrm{GHz}}$ for a pulsar at distance $d_{\textrm{kpc}}$ and where $r$
is the radius of the emission region (in meters). For characteristic values: $F \sim 1\:\mbox{Jy}$, $ d \sim 1\:\mbox{kpc}$ at $\nu \sim 1\:\mbox{GHz}$, 
this implies that the brightness temperature ranges from $10^{26}\:\mbox{K}$ to
$10^{30}\:\mbox{K}$, depending on whether the radius of the emission region is
taken to be the whole pulsar ($r = r_\ast \sim 10^4\:\mbox{m}$) or
only the polar cap ($r = r_{\textrm{pc}} = r_\ast (\Omega_\ast
r_\ast/c)^{-1/2}\sim 10^2\:\mbox{m}$ for $P_\ast = 2\pi/\Omega_\ast =
1\:\mbox{s}$).  
If the emission process were incoherent
this would imply the presence of energetic electrons (positrons) of
individual energies ${\cal E} \sim k_{\textrm{B}}\:T/m_{\textrm{e}} \sim 10^{10} - 10^{18}\:\mbox{GeV}$, which are difficult to achieve in view of the maximum available
voltage jump in a rotating magnetic star inside the light cylinder. 
Because of this, the mechanism responsible for the radio
emission is assumed to be some form of coherent action. The emission
can then be either an antenna or a maser process. 
If the radiation source is located on the open field lines above the
polar cap a maser emission process is most likely powered by a
high-energy electron/positron beam. Candidates for the emission
process are maser curvature emission \citep{lm1995}, relativistic plasma
emission \citep{mg1999}, normal and anomalous Doppler
instability \citep{PhD_lyutikov}, and linear acceleration emission \citep{m1978}. 

Despite the efforts in the investigations over the past three decades, the
mechanism responsible for the radio emission is still unknown.  
\\[\baselineskip]
The process studied here is a specific type of the free-electron
laser (FEL), a laboratory device which produces coherent radiation. In
a FEL, a beam of relativistic electrons, with velocity $v$, passes
through a periodic, electromagnetic field (called {\it wiggler}) and radiates at the resonance
frequency $\nu_{\textrm{res}} = \omega_{\textrm{res}}/2\pi = \pm(2\gamma^2 (\omega_{\textrm{w}} -
k_{\textrm{w}} v))/2\pi$, where $\omega_{\textrm{w}}$ and $k_{\textrm{w}}$ are the wave frequency and
the wavenumber respectively associated with the wavelength $\lambda_{\textrm{w}}$ of the wiggler. 
Under appropriate conditions the radiation will be enhanced by the
particles that follow, due to the bunching of the
particles. Previously mentioned linear acceleration emission
\citep{m1978} and coherent inverse Compton scattering \citep{srkl2002}
are both forms of this mechanism. They differ from the case here in the
sense that, in this study, particles will undergo a small transverse
displacement, whereas in the other two cases, the distortion is along
the path of the particles. Therefore, the solution proposed here is only
applicable for the region where the background magnetic field is
sufficient small, in comparison to the magnetic field amplitude of the
wiggler, i.e. high up in the magnetosphere. \\
Applying the FEL concept to the pulsar, we will investigate as
potential wigglers high-frequency
Alfv\'{e}n waves and other potential periodic structures in the pulsar
magnetosphere. Such waves might be
generated by a beam instability of the (remnant of the) primary beam
in the ambient secondary pair plasma, or by the inhomogeneity of the
pair plasma in which faster particles run into dense clumps of pairs
\citep{u1987,am1998,mgp2000}. The efficiency of the FEL interaction between these waves and the beam of primary/secondary particles is investigated by doing numerical simulations. 
\par 
In section~2 the basic concept of the free-electron laser mechanism
is presented in detail before it is applied to the pulsar
magnetosphere. A description of the simulation code is given is
section~3, and a summary of free-electron laser parameters is found in section~4, In section~5, we will present the results of the numerical
simulations of this process under pulsar magnetosphere conditions. The
conclusions and a discussion are given in section~6. 
\par
Throughout the paper, SI units and Cartesian coordinates, in which ${\bf x}$ is the position vector, with local magnetic field parallel to $\hat{z}$ are used. 

\section{The Model: A FEL in the pulsar magnetosphere}
\label{sec:model}
In this section the basic theory of FEL operation is explained. \\
Note that the FEL mechanism that we treat here is an antenne process in
which bunching occurs in space ("reactive"). 
\subsection{Free-electron laser theory}
\subsubsection{1-particle trajectory and resonance frequency}
The relativistic equation of motion for a particle in an
electromagnetic disturbance (wiggler), with electric field given by
${\bf E}_{\textrm{w}}$ and magnetic field by ${\bf B}_{\textrm{w}}$, is given by: 
    \begin{equation}
    \frac{\mbox{d}\gamma m \mbm{\beta} c}{\mbox{d}t} = q[{\bf E}_{\textrm{w}}
    + \mbm{\beta} c\times {\bf B}_{\textrm{w}}] \label{eq:reom} 
    \end{equation}
where the electromagnetic field of the wiggler is described by: 
    \begin{eqnarray}
    {\bf E}_{\textrm{w}}\rt & = & E_{\textrm{w}} \cos(k_{\textrm{w}} z - \omega_{\textrm{w}} t) \hat{x} \nonumber \\
    {\bf B}_{\textrm{w}}\rt & = & B_{\textrm{w}} \cos(k_{\textrm{w}} z - \omega_{\textrm{w}} t) \hat{y} \nonumber \\
    \frac{E_{\textrm{w}}}{B_{\textrm{w}}} & = & \frac{\omega_{\textrm{w}}}{k_{\textrm{w}}} = \beta_{\textrm{w}} c
    \label{eq:t_waves}
    \end{eqnarray} 
and $\beta_{\textrm{w}}$ denotes the wiggler phase velocity normalised to the velocity of light $c$. 
For a relativistic particle, with initial Lorentz factor $\gamma_0 = (1 - \beta_{z0})^{-1/2} \gg 1$, the first order
approximation of (\ref{eq:reom}), where we
neglect the particle's energy change, becomes: 
    \begin{equation}
    \frac{\mbox{d} \beta_x}{\mbox{d} t} =
    \frac{\sqrt{2}K}{\gamma_0}\omega_{\textrm{w}} \left( 1 -
    \frac{\beta_{z0}}{\beta_{\textrm{w}}}\right) \cos(k_{\textrm{w}} z - \omega_{\textrm{w}} t)
    \end{equation}
The solution of this equation is: 
    \begin{equation}
    \beta_x = \eta\frac{\sqrt{2}K}{\gamma} \sin (k_{\textrm{w}} z - \omega_{\textrm{w}} t)
    \label{eq:wigglebeta}
    \end{equation}
where $\eta = q/|q|$ and the strength of the wiggler is re-expressed in a dimensionless
parameter $K$, defined as:
    \begin{equation}
    K \equiv \left| \frac{q B_{\textrm{w}} \lambda_{\textrm{w}} }{\sqrt{8}\pi mc} \right|
    \label{eq:K}
    \end{equation}
Thus, the effect of the wiggler on the particle orbit is adding a
small sinusoidal motion in the direction of the wiggler's electric field, where the
periodicity of this motion for an observer in the laboratory frame is given by     \begin{equation}
    t_{\textrm{eff}} = \frac{\lambda_{\textrm{eff}}}{c} = \frac{\lambda_{\textrm{w}}}{ (\beta_{z0} - \beta_{\textrm{w}}) c}
    \label{eq:teff}
    \end{equation}

Up to the second order, the (angular) frequency of the radiation emitted by a particle moving in this disturbance is derived from $(\omega - {\bf k}\cdot{\bf v}) = \pm (\omega_{\textrm{w}} - k_{\textrm{w}} v_z)$: 
    \begin{equation}
    \omega_{\textrm{res}} = -\frac{2 \gamma^2(\omega_{\textrm{w}} - k_{\textrm{w}} v_z)}{1 + K^2}
    \label{eq:nu_obs}
    \end{equation} 
where we have used $\omega - {\bf k}\cdot{\bf v} \approx \omega(1 - \beta_z)$ for radiation beamed forward and $\beta^2_z = \beta^2 - \beta^2_x = 1 - \gamma^{-2} - K^2\gamma^{-2}$. 

\subsubsection{Collective effect: bunching}
The wiggler field causes a transverse motion of the particles (\ref{eq:wigglebeta}). In presence of
a radiation field ${\bf E}_{\textrm{rad}},\;{\bf B}_{\textrm{rad}} \propto \sin(k z -
\omega t)$, which is set up by particles radiating at the resonance frequency
(\ref{eq:nu_obs}), the particles that follow will move in a beat wave
of the wiggler field and the radiation field. This beat is often called
the ponderomotive wave, for its spatial energy density is highly
non-uniform. The corresponding ponderomotive force is ${\bf F}_{\textrm{p}} = q \beta_x c\hat{x} \times {\bf B_{\textrm{rad}}}$ and is therefore proportional to $\sin((k +
k_{\textrm{w}})z - (\omega + \omega_{\textrm{w}})t)$. The beat which is proportional to $\sin((k - k_{\textrm{w}}) z - (\omega - \omega_{\textrm{w}}) t)$ is 
superluminal, and, therefore, has no immediate particle-wave resonances. \\ 
This force acts on the beam in the $z$-direction and 
drives a longitudinal current $\delta J_z = q n v_{\textrm{p}}$, where $q$ is the
elementary charge of the beam particles, $n$ is the number density of the
particles and $v_{\textrm{p}}$ is the velocity which is induced by the motion 
in the ponderomotive wave. As a result, like ${\bf F}_{\textrm{p}}$, $\delta J_z$
is proportional to $\sin((k + k_{\textrm{w}})z - (\omega + \omega_{\textrm{w}})t)$ as
well. Related to $\delta J_z$ is a density modulation according to $q \partial \delta n/\partial t = - \mbm{\nabla}\cdot \delta J_z
\hat{z}$. This density modulation is observed as bunching of the
particles. Note that the bunches occur at the ponderomotive
wavelength 
    \begin{equation}
    \lambda_{\textrm{p}} = 2\pi k_{\textrm{p}}^{-1}
    \label{eq:pondwavelength}
    \end{equation}
where $k_{\textrm{p}} = k_{\textrm{w}} + k$. Due to
this bunching, the transverse current is in phase with the ambient
field ($\delta J_\perp = q\delta n {\bf v}_{\textrm{w}} \propto \sin(k z - \omega t)$), i.e. particles in the bunch radiate in phase and enhance the
ambient radiation field. 
 
\subsection{Background Magnetic Field}
The effect of a background magnetic field ${\bf B} = B_0 \hat{z}$ limits the
particle's motion perpendicular to the field. The equations of motion
for a particle now become: 
    \begin{align}
    \frac{\mbox{d} \gamma \beta_x}{\mbox{d}t} & =  \sqrt{2} K
    \omega_{\textrm{w}} \left(1 - \frac{\beta_z}{\beta_{\textrm{w}}}\right) \sin(k_{\textrm{w}} z -
    \omega_{\textrm{w}} t) + \Omega_{B_0} \beta_y \\
    \frac{\mbox{d}\gamma \beta_y}{\mbox{d}t} & =  \Omega_{B_0} \beta_x\\ 
    \frac{\mbox{d}\gamma \beta_z}{\mbox{d}t} & =  \sqrt{2}K
    \frac{\beta_x}{\beta_{\textrm{w}}} \omega_{\textrm{w}}
    \end{align}
where $K$ is defined by equation (\ref{eq:K}) and $\Omega_B = |q|B/m$ is
the non-relativistic gyrofrequency for a particle (mass $m$, charge
$q$) in a magnetic field with strength $B$. For zero background
magnetic field, the equations reduce to (\ref{eq:reom}). 
In the presence of a uniform background
magnetic field, the first order solution for the particle's velocity
components (the $x$- and $y$-component) are coupled to each other
through $\Omega_{B_0}$. The particle is no longer free
to resonate with the wiggler, but is bounded by the magnetic
field. This restriction limits the particle's ability to resonate and, therefore, also the FEL laser action. 
\subsection{Application in the pulsar magnetosphere}
\begin{figure}
    \includegraphics[width=0.45\textwidth]{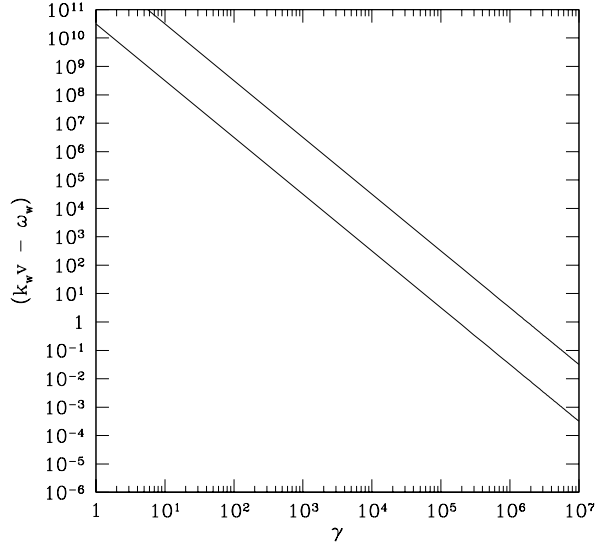}
    \caption{Values for which $\omega_{\textrm{w}}$, $k_{\textrm{w}}$ and $\gamma$ give rise
      to radio emission between $100\:\mbox{MHz}$ (lower solid line)
      and $10\:\mbox{GHz}$ (upper) according to the resonance condition
      (\ref{eq:nu_obs}) with $K = 10$. On the vertical axis is plotted $k_{\textrm{w}} v - \omega_{\textrm{w}}$, because we assume the particle's velocity to be larger than the wiggler's phase velocity.  } 
    \label{fig:parameterplot}
    \end{figure}

Our aim is to apply the mechanism to the pulsar magnetosphere, while
making as few assumptions as possible. As is common to most pulsar models, we assume that a large electric field is set up along the open field lines
above the polar caps due to the fast rotation of the magnetised
neutron star. This electric field pulls out and accelerates electrons, which, in the presence of a strong magnetic field, are in
the lowest Landau level and thus move along the magnetic field
lines. 
At some altitude above the pulsar surface - between a few polar cap
radii and a few stellar radii - a dense plasma of electron-positron
pairs is produced, either from curvature radiation of the primaries in
the strong magnetic field or from their inverse Compton radiation. Very
likely, the entire process is only stationary in an average sense but
highly variable in space and time on small scales with inhomogeneous
distributions of pair plasma intermingled with primary beams. 
\\[\baselineskip]
In our model, we investigate radiation from a mono-energetic beam of
electrons in the presence of a subluminal transverse electromagnetic wiggler.
\\[\baselineskip]
In theory, radiation of any frequency can be obtained by tuning
$\omega_{\textrm{w}}$, ${\bf k}_{\textrm{w}}$ and $\gamma$. For radiation at frequencies 
$\nu_{\textrm{res}} = \omega_{\textrm{res}}/2\pi$ in the range of $10^{8}$ and
$10^{10}\;\mbox{Hz}$, and parallel wave vectors ${\bf k}_{\textrm{w}} \parallel {\bf
  v}$, the required Lorentz factors range from unity up to $10^7$ (Fig.~\ref{fig:parameterplot}). 

Near the stellar surface, particles move one-dimensionally along the
magnetic field lines due to extremely fast synchrotron losses in the strong magnetic field. The quantity $p_\perp^2/B_0$ is not invariant anymore when the timescale at which the magnetic
field changes, is small compared to one gyration period in the background
magnetic field, i.e.:
    \begin{equation}
    \frac{B_0}{\mbox{d}B_{\textrm{w}}/\mbox{d}t} \ll \frac{2 \pi \gamma}{\Omega_{B_0}} 
    \end{equation} 
and in our case, using (\ref{eq:t_waves}):  
    \begin{equation}
    B_0  \ll  2 \pi\sqrt{\frac{\gamma mc}{q}\frac{B_{\textrm{w}}(\beta_{z0} - \beta_{\textrm{w}})}{\lambda_{\textrm{w}}}} \label{eq:b0maximum}
    \end{equation}
So, if the last inequality is satisfied, a particle moving in the
background magnetic field acquires momentum transverse to the magnetic field. \\ 
Next to consider is the synchrotron loss time, which needs to be larger
than the FEL timescale when the latter process causes the pulsar radio
emission. 
The synchrotron loss time $\tau_{1/2}$, the time within which the particle
looses half of its initial energy $E = \gamma m c^2 $, is defined as $\tau_{1/2} \equiv E/(2P)$, where
$P = \gamma^2 \beta_\perp^2 \Omega_{B_0}^2/(6 \pi \epsilon_0 c)$ is the
power of synchrotron radiation emitted by a charged particle moving in
a background magnetic field $B_0$ and $\beta_\perp$ is the component of the normalised velocity perpendicular to $B_0$. By using the upper limit for $B_0$ (\ref{eq:b0maximum}), we find for $\tau_{1/2}$
    \begin{equation}
    \tau_{1/2} \gg \frac{1}{2}\frac{3 \epsilon_0 mc}{2 \pi q^2}\frac{\lambda^2_{\textrm{w}}}{(\beta_{z0} - \beta_{\textrm{w}})^2}\frac{1}{K^2 \gamma}
    \label{eq:tauhalf} 
    \end{equation}
where we used $\beta_\perp \approx K/\gamma$ from~(\ref{eq:wigglebeta}). \\
On timescales much smaller than $\tau_{1/2}$,
a change of the particle energy due to synchrotron radiation can be
neglected. \\
In our calculations, we used parameters such that firstly, the resonance is in the radio regime (\ref{eq:nu_obs}), secondly, the inequality (\ref{eq:b0maximum}) is satisfied so that particles can (and do) acquire transverse momentum, and finally, the synchrotron emission can be neglected (\ref{eq:tauhalf}). 

\section{Numerical Simulations}
This section describes how the ingredients for the FEL in the
pair plasma of a pulsar magnetosphere are represented in our numerical
simulations. The Coulomb interactions between the particles in the beam are
neglected, due to the large Lorentz factors. This is because, although the
beam particles generate an electric field radially outward $E_r$, due to
their relativistic speeds, the generated magnetic field (in the
azimuthal direction) reduces the electric field by a factor of
$1 - \beta^2$, where $\beta$ is the particles' speed. This results in a
reduced radial Lorentz force $F_r = q E_r/\gamma^2$. \\
Furthermore, the role of the pairplasma is mainly in providing the wiggler.
Therefore, we model the presence of the wiggler by equations~(\ref{eq:t_waves}) rather than generating it in a consistent way. \\
First, we give a brief introduction to the code. Then, the parameters for
each ingredient in the case of $N$-particles simulations are given. 
\subsection{Code General Particle Tracer (GPT)}
\begin{figure}
    \includegraphics[width=0.4\textwidth]{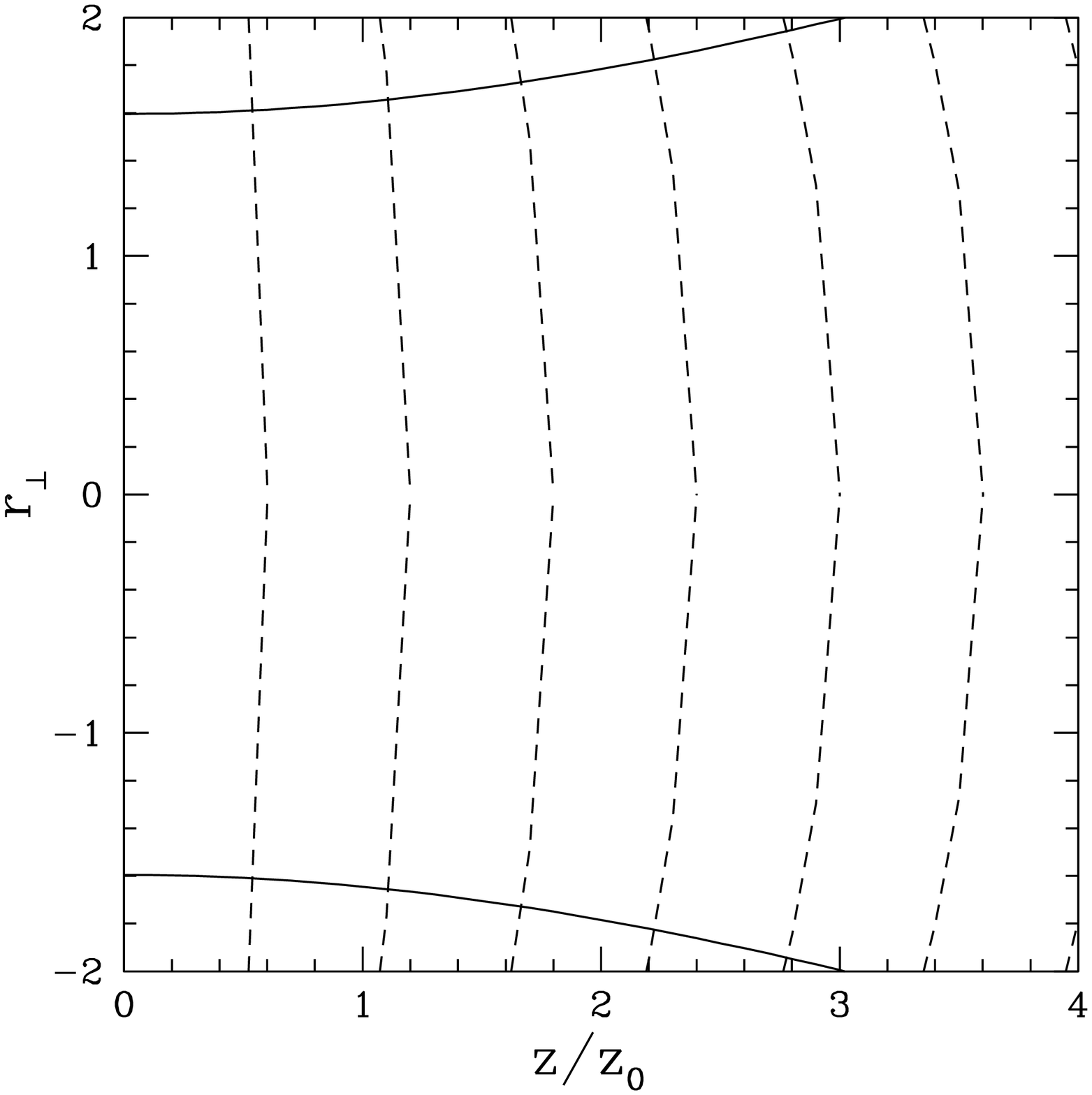}    
    \includegraphics[width=0.4\textwidth]{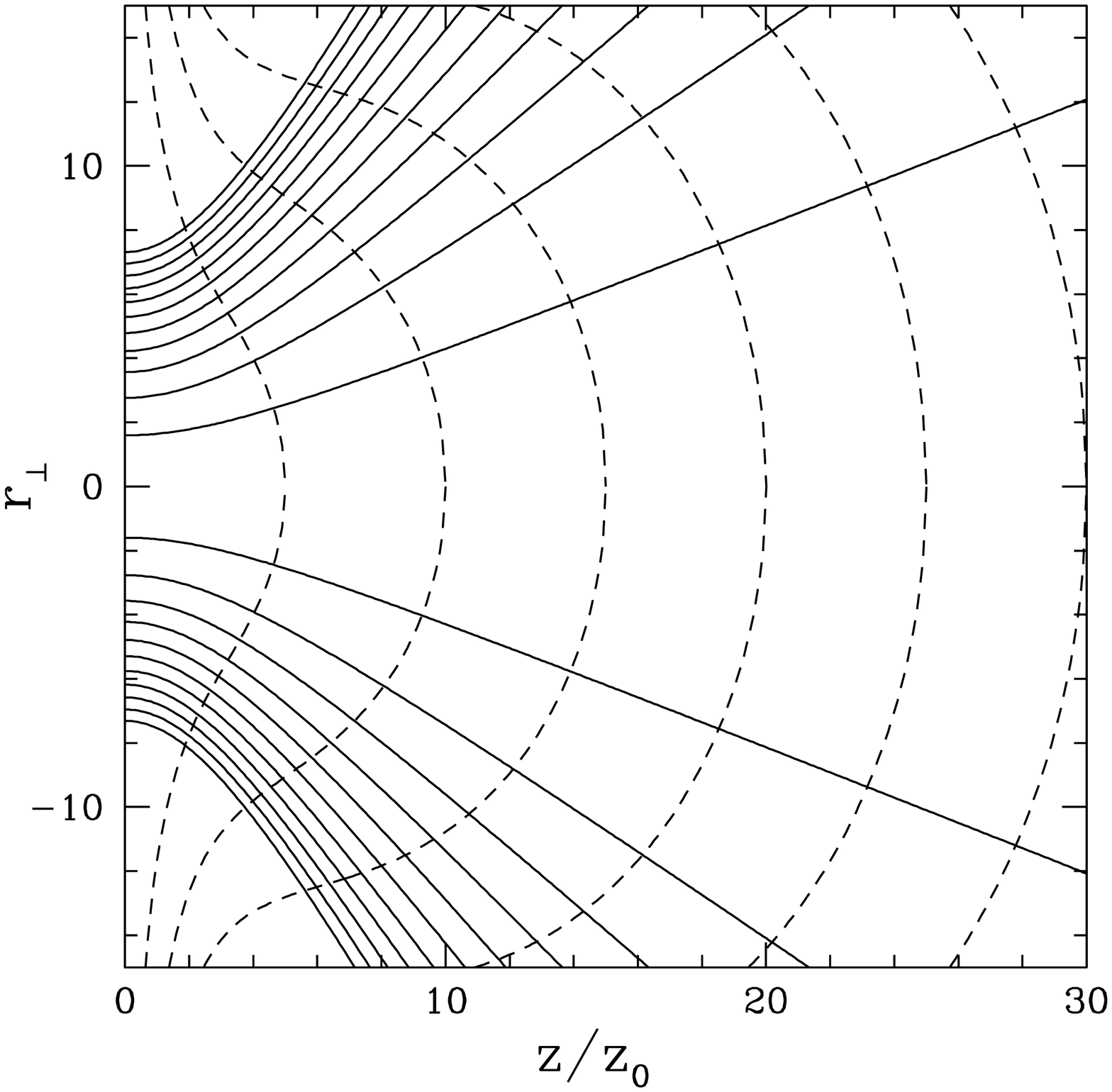}
    \caption{Wavefronts (dotted lines) and level curves at $e^{-n}$ of the maximum
    amplitude of the Gaussian mode (solid lines, $n = 1, 2, ..$);
    left: $near$-field, right: $far$-field. $z$ is in units of $z_0$
    and $r_\perp$ in units of $w_0$. }
    \label{fig:nearfar}
    \end{figure}

We will investigate radiation from a relativistic electron beam travelling through a
wiggler, and the formation of bunches is tested by doing numerical
simulations with a code called General Particle Tracer
(GPT)\footnote{see http://www.pulsar.nl/ for a description of the
  code and other publications with this code}. \\
This code solves the equation of motion of each 
(macro)particle in the time domain numerically. For each
macroparticle, labelled $i$, the differential equations 
    \begin{eqnarray}
    \frac{\mbox{d}\gamma_i \mbm{\beta}_i}{\mbox{d}t} & = & \frac{q_i}{m_ic}[{\bf E}\rt + {\bf v}_i(t) \times {\bf B}\rt] \nonumber \\
    \frac{\mbox{d}{\bf x}_i}{\mbox{d}t} & = & {\bf v}_i = \frac{\gamma_i\mbm{\beta}_i c}{\sqrt{\gamma_i^2\mbm{\beta}_i^2 + 1}} \label{eq:eoms} 
    \end{eqnarray}
are solved numerically with a fifth-order embedded Runge-Kutta ODE
solver \citet{bookNumericalRecipes}. Here $\gamma = (1 - \mbm{\beta}^2)^{-1/2}$ is the Lorentz factor, $\mbm{\beta} = {\bf v}/c$, and ${\bf E}\rt$, ${\bf B}\rt$ are the
electromagnetic fields in which the particle is moving at position
${\bf r}$ and time $t$. Because a
macroparticle represents $N_q$ particles of elementary mass $m_{\textrm{e}}$ and charge $-e$, the
fraction $q/m$ in the equation of motion is the same as in the case of
single electrons.  
The electric field consist of two parts, ${\bf E}\rt = {\bf E}_{\textrm{w}}\rt + {\bf E}_{\textrm{rad}}\rt$, where ${\bf E}_{\textrm{w}}\rt$ is the
wiggler field and ${\bf E}_{\textrm{rad}}\rt$ is the radiation field,
of which the development is studied in time; the same holds for ${\bf
  B}\rt$. The space charge effects of the
particles in the beam are neglected. 
The initial
conditions of the particles (i.e. number of charged particles, mass, charge,
initial distributions in coordinate and velocity space) as well as the
spatio-temporal behaviour (including initial amplitude and phase) of
the wiggler, are user-specified.  
\\[\baselineskip]
To find the radiation in the simulations, a set of differential equations for the radiated energy are solved, additional to the differential
equations~(\ref{eq:eoms}). To derive this set of differential
equations, we first note that there are two essential differences
between a FEL in the lab and our astrophysical application: firstly,
there are no reflective mirrors in the neutron star magnetosphere. For
a FEL to operate under such conditions, it should be a high-gain,
single-pass process. Secondly, there are no side walls bounding the
FEL cavity in the magnetosphere. This is accommodated in our pulsar
study by using Gaussian modes of which the field decreases in a
gaussian manner to zero away from the axis. \\
The radiation electric field is splitted into a set of
Gaussian modes (see Appendix A): 
    \begin{eqnarray}
    {\bf E}_{\textrm{rad}}\rt & = & \sum_j E_j\rt \hat{x} \\
    E_j\rt & = &   A_j T_j \cos(\theta_j)  \label{eq:builtup} 
    \end{eqnarray}
where 
    \begin{eqnarray}
    T_j & = & \frac{w_0}{w_j(z)}e^{-r_\perp^2/w_j^2(z)} \\
    \theta_j & = & \omega_j t - k_j z - \frac{k_j r_\perp^2 z}{2(z^2 +
    z^2_{0,j})} - \arctan\left(\frac{z}{z_{0,j}}\right) + \xi_j, 
    \end{eqnarray}
where we have chosen cylindrical coordinates ($z$, $r_\perp = (x^2 +
y^2)^{1/2}$), and the polarisation
is chosen to lie in the $x$-direction, in agreement with the
polarisation of the wiggler which is expected to generate radiation
polarised in the $x,z$-plane. Note that, there should be a small
component of electric field in the $z$-direction as well. But since this
is negligible, as compared to the $x$-component of the Gaussian mode, its
interaction with the particles is neglected in the calculations. \\
Further, $w_j(z) = w_0 \sqrt{1 + z^2/z_{0,j}^2}$ is the {\it waist},
$w_0$ is the waist at $z=0$, and  $z_{0,j} = (1/2)k_jw_0^2$ is the
characteristic distance where the wave starts to diverge, $A_j$ is the
amplitude of this wave at wavevector $k_j$, $\omega_j = k_jc$ is
the corresponding angular wavefrequency, and $\xi_i$ is an arbitrary phase. \\
Note that Gaussian modes propagate initially only in one direction
and, therefore, they are suitable to describe narrow beams of pulsar
radio emission. Figure~{\ref{fig:nearfar}} shows the wavefront of a Gaussian mode
(dotted line); the solid lines are level curves at $\mbox{e}^{-1,-2,..}$ of the
maximum amplitude (left: near-field, right: far-field). 
As the figure shows, the wavefront of a Gaussian mode changes from
planar to spherical. The asymptotic opening angle $\phi$ of this
wave is given by: 
    \begin{equation}
    \tan{\phi} =  \sqrt{\frac{\lambda}{2 \pi z_0}}
    \label{eq:openangle}
    \end{equation}
In the GPT, the user specifies the minimum and the maximum frequency of
the Gaussian modes and the number of modes in this range. The frequency intervals are equally spaced (given by $\Delta N_m$) and each is
represented by a Gaussian mode with wavevector $k$. The spectrum in
this range is represented by the sum of these modes
(equation~(\ref{eq:builtup})). When the radiation of the pulsar is assumed to be due to a FEL, then this set of Gaussian modes describes the radiation properly. \\
To continue the derivation of additional differential equations, we
use conservation of energy: the energy gain/loss per unit time of the $i$th macro-particle with
 charge $Q_i = N_q q_i$ due to the interaction with the $j$th electric field
 component is given by: $n q_i\:{\bf v}_i \cdot {\bf E}_j$, 
 which is balanced by the rate of change of the electromagnetic
 fields. The total power radiated $P = \mbox{d}W_{\textrm{light}}/\mbox{d}t$ is obtained from: 
    \begin{equation}
    \sum_{i,j}N_q q_i\:{\bf v}_i \cdot {\bf E}_j + \frac{\mbox{d}W_{\textrm{light}}}{\mbox{d}t} =0
    \end{equation}
where the summation is over all $N$ particles and $N_m$ Gaussian modes. \\
Instead of calculating the change in amplitude and phase of each mode
directly, GPT rewrites the amplitudes and phases into two other independent variables $m_j$ and $n_j$: 
    \begin{eqnarray}
    A_j & = & \sqrt{m_j^2 + n_j^2}/\Delta N_m\nonumber \\
    \phi_j & = &\arctan(n_j/m_j)
    \label{eq:akphik}
    \end{eqnarray} 
Inserting equations~(\ref{eq:akphik}) into the conservation of energy
gives two differential equations for $m_j$ and $n_j$ ($j$ = 1.. $N_m$): 
    \begin{eqnarray}
    \frac{\mbox{d}{m_j}}{\mbox{d}t} & = & -\sum_i \frac{N_q q_i \Delta N_m}{\pi w_0^2 \epsilon_0 L} v_{x,i} T_j \cos(\theta_j) \nonumber \\
    \frac{\mbox{d}{n_j}}{\mbox{d}t} & = & \sum_i \frac{N_q q_i \Delta N_m}{\pi w_0^2 \epsilon_0 L} v_{x,i} T_j \cos(\theta_j) 
    \label{eq:diffuv}
    \end{eqnarray}
where $L$ is the cavity length. \\
A factor $\Delta N_m$ enters Eqs.~(\ref{eq:akphik}) and
(\ref{eq:diffuv}) because one mode represents a frequency interval. This
gives the correct calculation of $m_j$ and $n_j$, which is dependent on
$\Delta N_m$, and ensures that the resulting spectrum, represented by $A_j$'s,
is independent on the number of modes used. 
\\
The wiggler part of the electromagnetic field is assumed to have a
time-independent amplitude during passage of the beam,  
and is not included here, for we assume that the change in the wigglers' energy is small on the considered timescales. 
\\[\baselineskip]
Summarising, the differential equations solved at
time $t$ by GPT consist of the $2N$ equations: 
    \begin{eqnarray}
    \frac{\mbox{d}\gamma_i \mbm{\beta}_i}{\mbox{d}t} & = & \frac{q_i}{m_ic}[{\bf E}\rt + {\bf v}_i(t) \times {\bf B}\rt] \nonumber \\
    \frac{\mbox{d}{\bf x}_i}{\mbox{d}t} & = & {\bf v}_i =
    \frac{\gamma_i\mbm{\beta}_i c}{\sqrt{\gamma_i^2\mbm{\beta}_i^2} +
    1} \nonumber 
    \end{eqnarray}
for the $N$ charged particles' phase space positions, and the set of $2N_m$ equations: 
    \begin{eqnarray}
    \frac{\mbox{d}{m_j}}{\mbox{d}t} & = & -\sum_i \frac{N_q q_i \Delta N_m}{\pi w_0^2 \epsilon_0 L} v_{x,i} T_k \cos(\theta_j) \nonumber \\
    \frac{\mbox{d}{n_j}}{\mbox{d}t} & = & \sum_i \frac{N_q q_i \Delta N_m}{\pi w_0^2 \epsilon_0 L} v_{x,i} T_k \cos(\theta_j) \nonumber 
    \end{eqnarray}
for the total number of modes $N_m$ included in the simulation, to
calculate the amplitude and phase at {\it each} wavenumber in the
specified range during the interaction with the particles. As a result
we obtain the spectrum of the radiation in time. 
\\[\baselineskip]
To retrieve the pulse in the time domain, both the electric field and the
power are Fourier transformed from the frequency to the time domain, and Parseval's
theorem is applied to the results to ensure a correct transformation between time and frequency domain.
\\ Note that the results obtained with this code are independent of
the number of time outputs, the number of macroparticles and the number of modes. 

\section{Physical Parameters for Numerical Simulations} 
Except in the second last run of the simulations, we used a beam of
200 macroparticles with a total charge of $-6 \cdot
10^{-4}\:\mbox{C}$. These are uniformly distributed in a cylinder with
radius of one metre and length of three ponderomotive wavelengths (\ref{eq:pondwavelength}), which varies from run to run. The corresponding
electron number density is $1.59 \cdot 10^{16}\:\mbox{m}^{-3}$. This is 0.23
times the Goldreich-Julian density at the stellar surface, given by
$n_{\textrm{GJ}}(r_\ast) = 2 \epsilon_0 \mbm{\Omega}\cdot{\bf B}_\ast/q$, for a
reference pulsar with $\Omega = 2 \pi/ P$, where $P = 1\:\mbox{s}$ and ${\bf
  B}_\ast = 10^8\:\mbox{T}$.  
Except in the last run, each macro-particle's initial Lorentz factor is $\gamma = 1000$. 
\\[\baselineskip]
The wiggler is an electromagnetic disturbance propagating in the
$z$-direction. In the pulsar magnetosphere, the scale at which
the physical parameters change, e.g. background magnetic field, is of
the order of $z/r_\ast$ ($\geq 10\:\mbox{km}$). Therefore, the
interaction between the beam and the wiggler is simulated until the
beam particles reach this distance, so $t_{\textrm{end}} \approx
3.3\cdot 10^{-5}\:\mbox{s}$. The parameters for the wiggler are given in
Table~{\ref{tabel:inout}} (all labelled with ``$\mbox{w}$''). The values for the
magnetic induction and angular frequency are obtained from (\ref{eq:t_waves}). \\
In the first run, the wiggler's phase speed is $0.9c$. The amplitude
of the magnetic induction $B_{\textrm{w}}$ is $3 \cdot 10^{-3}\:\mbox{T}$. The wavelength $\lambda_{\textrm{w}}$ is 50 metres. These values are chosen such that the
resonance is at radio frequencies, and the strength of the magnetic induction is
chosen such that the induced transverse velocity is small compared to
the initial axial velocity ($K/\gamma \ll 1$ in
Equation~(\ref{eq:wigglebeta})). \\
In the subsequent runs we investigate the effects of varying each
of these parameters as compared to the results of the first run. In run~2, the wiggler strength $K$ is increased from five to twenty, in steps of $K=5$ (correspondingly the magnetic induction is increased in steps of $B_{\textrm{w}} = 1.5 \cdot 10^{-3}\:\mbox{T}$). In the third
run, the wavelength of the wiggler takes the values 25, 40, 50 and 100 m. In run~4, the velocity of
the wave is changed from 0.2 to 0.9. 
\\[\baselineskip]
In run~5, we study the effect of a non-zero background magnetic
field. We took the most simple case is which the field in uniform (in
$z$) and has a background magnetic field strength of $B_0 = 10^{-3}$, $10^{-2}$, $2.5 \cdot 10^{-2}$, $5 \cdot 10^{-2}$ T. 
\\[\baselineskip]
In run~6 as we change the number density $n$ of the electron
beam. In
terms of the Goldreich-Julian density, the fractional density ranges from
0.4 to 0.1. 
\\[\baselineskip]
Finally, in run~7, we lower the Lorentz factor of the beam particles (the resonance frequency depends on this as $\omega_{\textrm{res}} \propto \gamma^2$). The Lorentz factors are chosen: $\gamma = 1000$, $500$, $250$, $100$. 

\section{Results}

\begin{table*}[t!]
\begin{scriptsize}
\begin{tabular}{c|cccccc|cccc|cccc}
\hline
{\bf Run} & \multicolumn{6}{l}{{\bf Input parameters}} & \multicolumn{4}{l}{{\bf Output}} & \multicolumn{4}{l}{{\bf Energy densities}} \\
\hline
 & $\gamma$& $n$ & $K$ & $\lambda_{\textrm{w}}$ & $\beta_{\textrm{w}}$ & $B_0$ & $\nu_{\textrm{cen}}$ & $\Delta \nu$ & $P_{\textrm{max}}$ & $T_{\textrm{b}}$ & $\epsilon_{\textrm{b}}$ & $\epsilon_{\textrm{w}}$ & $\epsilon_{\textrm{b0}}$ & $\epsilon_{\textrm{p}}$ \\
 & & ($n_{\textrm{GJ}\ast}$) & & (m) & & (T) & (GHz) & (GHz) & ($10^{13}\:\mbox{W}$) & ($10^{30}\:\mbox{K}$) & ($10^6$ Jm$^{-3}$) & (Jm$^{-3}$) &  (Jm$^{-3}$) &  ($10^2$ Jm$^{-3}$) \\ 
\hline
1 & 1000 & 0.23 & 10 & 50 & 0.9 & 0 & 11 & 2.42 & 9.09 & 1.02 & 1.30 & 8.15 & 0 & 8.25  \\ 
2 & - & - & 5 & - & - & - & 42.8 & 9.72 &36.6 &1.04 &- &2.04 &- &34.5 \\
  & - & - & 10 & - &- &- & 11 &2.42 &9.09 &1.02 &- &8.15 &- &8.25 \\  
  & - & - & 15 & - &- &- &4.96 & 1.02 &3.67 &1.16& - &8.35& - &3.46 \\
  & - & - & 20 & - &- &- &2.84 & 0.53 &1.61 &0.99 &- &32.62& - &1.50 \\
3 & - & - & - & 25 & - & - & 22.2 & 3.8 & 9.52 &  1.44 & - & 32.6 & - & 9.00 \\
& - & - & - & 40 & - & - & 13.8 & 2.86 & 9.42 & 1.02 & - & 10.07 & - & 8.89 \\
& - & - & - &50 & - & - & 11 & 2.42 & 9.09& 1.34&  - & 8.15 & - & 8.25 \\
  & - & - & - & 100 & - & - & 5.57 & 1.28 & 7.02 & 1.70 & - & 2.03 & - & 6.63 \\
4 & - & - & - & - & 0.2 & - & 88.5 & 14.1 & 9.59&  0.215&  - &94.9& -& 9.00 \\
  & - & - & - & - & 0.5 & - & 55.7 & 8.12 & 9.54 & 0.37 & - & 18.2 & - & 9.00 \\
  & - & - & - & - & 0.8 & - & 22.2 & 3.79 &9.39 &0.79 &- &9.35 &- &8.86 \\
  & - & - & - & - & 0.9 & - & 11 & 2.42 &9.09 &1.02 &- &8.15 &- &8.25 \\
5 & - & - & - & - & - & $10^{-3}$ & 10.9 & 2.43& 9.46 &7.4 &- &- &0.4 &8.85 \\
  & - & - & - & - & - & $0.01$ & 2.18 & 0.61 &2.7 &1.36 &- &- &40 &2.55 \\
  & - & - & - & - & - & $0.025$ & n & n & 0.455 & n & - & - & $2.48 \cdot 10^2$& $42.8 \cdot 10^{-2}$\\ 
  & - & - & - & - & - & $0.05$ & n & n & $6.36 \cdot 10^{-4}$ & n & - & -& $9.95 \cdot 10^{2}$ & $4.0 \cdot 10^{-4}$ \\ 
6  & - & $0.1$ & - & - & - & - & 11.37 & 1.53 & 1.79 & 0.38 & 2.30 & -& -& $1.7$\\ 
  & - & $0.2$ & - & - & - & - & 11.1 & 2.22 & 7.2 & 1.07 & 1.76 & - & - & $6.8$\\ 
  & - & $0.3$ & - & - & - & - & 10.9 & 3.01& 16.5 & 1.73 & 1.15 & - & - & $15.5$\\ 
  & - & $0.4$ & - & - & - & - & 10.7 & 3.40 & 29.9 & 2.8 &0.575& - &- & $28.2$ \\ 
7 & 1000 & -  & - & - & - & - & 11 & 2.42 &9.09 &1.24 & 1.3 &- &- &8.24 \\
  & 500 & -  & - & - & - & - & 2.72 &0.80 &1.77 &0.72 &0.16 &- &- &1.66 \\
  & 250 & -  & - & - & - & - & 0.69 &0.16 &0.112 &0.23& 0.02 &- &-&0.10 \\
  & 100 & -  & - & - & - & - & 0.09 & 0.006 & $2.7\cdot 10^{-4}$ &$4.6\cdot 10^{-3}$ & $1.36\cdot 10^{-3}$&- & - & $2.5 \cdot 10^{-4}$ \\
\hline 
\end{tabular}
\end{scriptsize}
\caption{Input parameters and results for all simulation runs. Input
  parameters are: electron initial Lorentz factor $\gamma_0$, number density in the beam $n$ in terms of Goldreich-Julian density at the pulsar surface $n_{\textrm{GJ}\ast}$, dimensionless
  wiggler strength $K$, wiggler wavelength $\lambda_{\textrm{w}}$,
dimensionless wiggler phase speed $\beta_{\textrm{w}}$ and 
  (uniform) background magnetic field $B_0$. Results from the numerical
  simulations are: central frequency $\nu_{\textrm{cen}}$, FWHM
  bandwidth of the radiation $\Delta \nu$, peak power of the light
  pulse $P_{\textrm{max}}$, and brightness temperature corresponding to the
  peak power $T_{\textrm{b}}$. The last four entries are the energy density of
  each ingredient in the simulation: beam $\epsilon_{\textrm{b}}$, 
  wiggler $\epsilon_{\textrm{w}}$, background magnetic field $\epsilon_{\textrm{b0}}$
  and pulse $\epsilon_{\textrm{p}}$. Note that every time ``-'' appears, it
  means the entry has the value as in Run~1. In Run~5, {\it n} appears for $\nu_{\textrm{res}}$, $\Delta \nu$ and $T_{\textrm{b}}$, which means, that for those cases, the spectra are flat, no peaks are seen, and also the brightness temperatures which depend on the bandwidths are not calculated. }
\label{tabel:inout}
\end{table*}

Figs.~{\ref{fig:bunching}}(Run~1) to {\ref{fig:n200gamma_tgam}}(Run~7) and
Table~{\ref{tabel:inout}} show the end results of all runs, i.e. at $t
= 3.3 \cdot 10^{-5}\:\mbox{s}$. Rather than discussing the runs
sequentially, we present the systematic trends of our computations,
using the figures as illustrations. Run~5, where a uniform
background field is included, is presented separately.  

\subsection{Bunching}
\begin{figure}[t!]
    \includegraphics[width=0.45\textwidth]{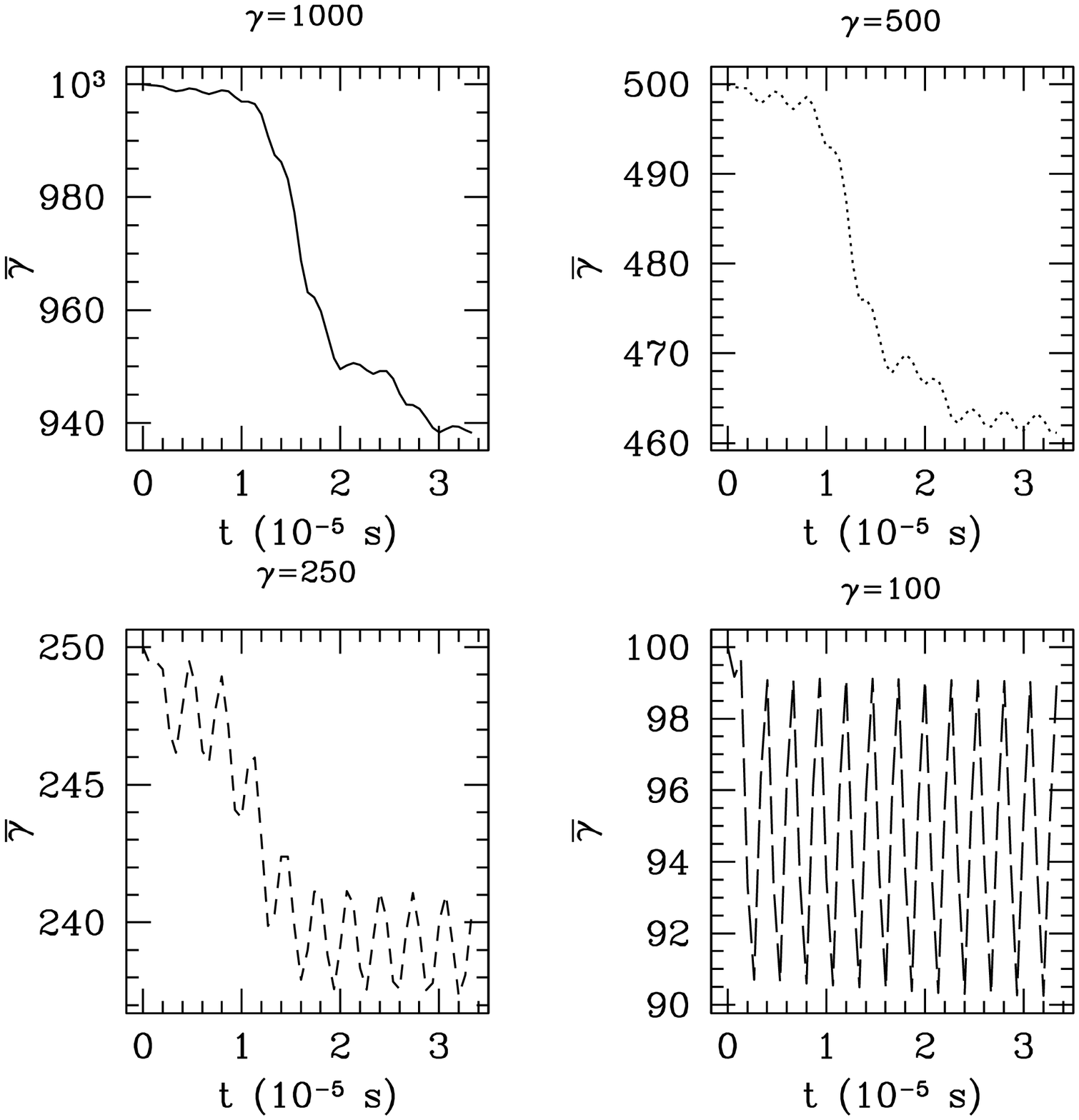}
    \mycaption c{Run~7: The average Lorentz factor versus the average distance $z$ of the particles in simulations with changing Lorentz factor. } 
    \label{fig:n200gamma_tgam} 
    \end{figure}

We found bunching of particles during their interaction with the
wiggler in a number of cases. Run~1 clearly shows this bunching (Fig.~{\ref{fig:bunching}}). Here, the particles are plotted in
($z$,$x$)-projection at different stages of their interaction: 
bunch formation ($t = 1.33 \cdot 10^{-5}\:\mbox{s}$), `de-bunching' ($t = 2.0
\cdot 10^{-5}\:\mbox{s}$) and `re-bunching' ($t = 2.67 \cdot
10^{-5}\:\mbox{s}$). Note that the distance between bunches is $\lambda_{\textrm{p}}$ as expected. \\ 
An alternative illustration of the formation of bunches is found in
Fig.~{\ref{fig:avggamma}}, where the average Lorentz factor per
macro-particle $\bar{\gamma}$ is plotted. Evidently, the formation of
bunches corresponds to a steep drop of $\bar{\gamma}$ (e.g. Fig.~{\ref{fig:avggamma}}, the first bunch formation starts from $t \simeq 1.0\cdot 10^{-5}\:\mbox{s}$ to $1.8 \cdot 10^{-5}\:\mbox{s}$, and the second formation from $2.4 \cdot 10^{-5}\:\mbox{s}$ to $2.8 \cdot 10^{-5}\:\mbox{s}$).  \\
We notice that both the bunch duration and the beam energy loss is larger for the first bunch formation. Saturation is reached due to an increased velocity spread. Therefore, for the second bunch formation, the beam starts with a certain spread in velocity, and it reaches saturation quicker than in the initial
bunching. \\
For the beam number density of $n = 0.23\:n_{\textrm{GJ}\ast}$, we find
a beam energy loss to radiation of $\sim 5 \%$, after the first time bunching. This number is only affected by $n$. The number of particles in {\it each} bunch scales with $n$, as does $\Delta \bar{\gamma}$ (Run~6, Fig.~{\ref{fig:n200nfac4_tgam}}), which agrees with coherent losses. \\
As to the starting time of the first bunching, we find the following: for the same beam and different wiggler parameters ($\beta_{\textrm{w}}$ and $\lambda_{\textrm{w}}$), about ten oscillations of the particles in the wave
{\it as seen by the observer} are needed before they start to bunch (Run~3, Figs.~{\ref{fig:N200lamu_tgam}} and Run~4, Fig.~{\ref{fig:N200vfac_tgam}}). For the same wiggler, increasing the beam density or decreasing the Lorentz factor give rise to earlier occurence of bunching (Run~6, Figs.~{\ref{fig:n200nfac4_tgam}} and Run~7, Fig.~{\ref{fig:n200gamma_tgam}}). This suggests that the bunching sets in when the (incoherent) radiation reaches a certain level, because the ponderomotive force is dependent on the radiation field.  \\

As for the wiggler strength $K$, it determines the maximum beam energy
loss during bunching. In Fig.~{\ref{fig:N200K_tgam}}(Run~2) are shown ($\bar{\gamma}$, $t$)-plots for $K = 5, 10, 15, 20$. We have also run cases for $K \leq 1$. The maximum energy loss occurs when $K = 1$. 

\subsection{Coherent radiation}
That the radiation is coherent is demonstrated by the power of the radiation
pulse (Fig.~\ref{fig:power}). At the top is presented the simulation run for $N = 200$
macro-particles and at the bottom for $N = 2$ macro-particles. The
maximum pulse power for $N = 200$ is $P_{200} = 9.09 \cdot 
10^{13}\:\mbox{W}$ and for $N = 2$, $P_{2} = 9.75 \cdot
10^9\:\mbox{W}$. If the radiation were incoherent emission from 200
macro-particles, we would expect $P \propto N$, where $N$ is the
number of particles and $(P_{200}/P_2)_{\textrm{inc}} = 100$. However,
the simulation shows that $(P_{200}/P_2) \simeq 9 \cdot 10^3$ , and
thus $P \propto N^2$, demonstrating that the radiation is coherent. \\
Note that only the maximum of the first peak satisfies $P_{200} \propto N^2$. This is not the case for the second peak in Fig.~{\ref{fig:power}}, where now $P_{200} \approx 2 \cdot 10^{13}\:\mbox{W}$, which is less than the coherent case but much more than if it were incoherent.

\subsubsection{Spectrum}
\begin{figure}[t!]
    \includegraphics[width=0.45\textwidth]{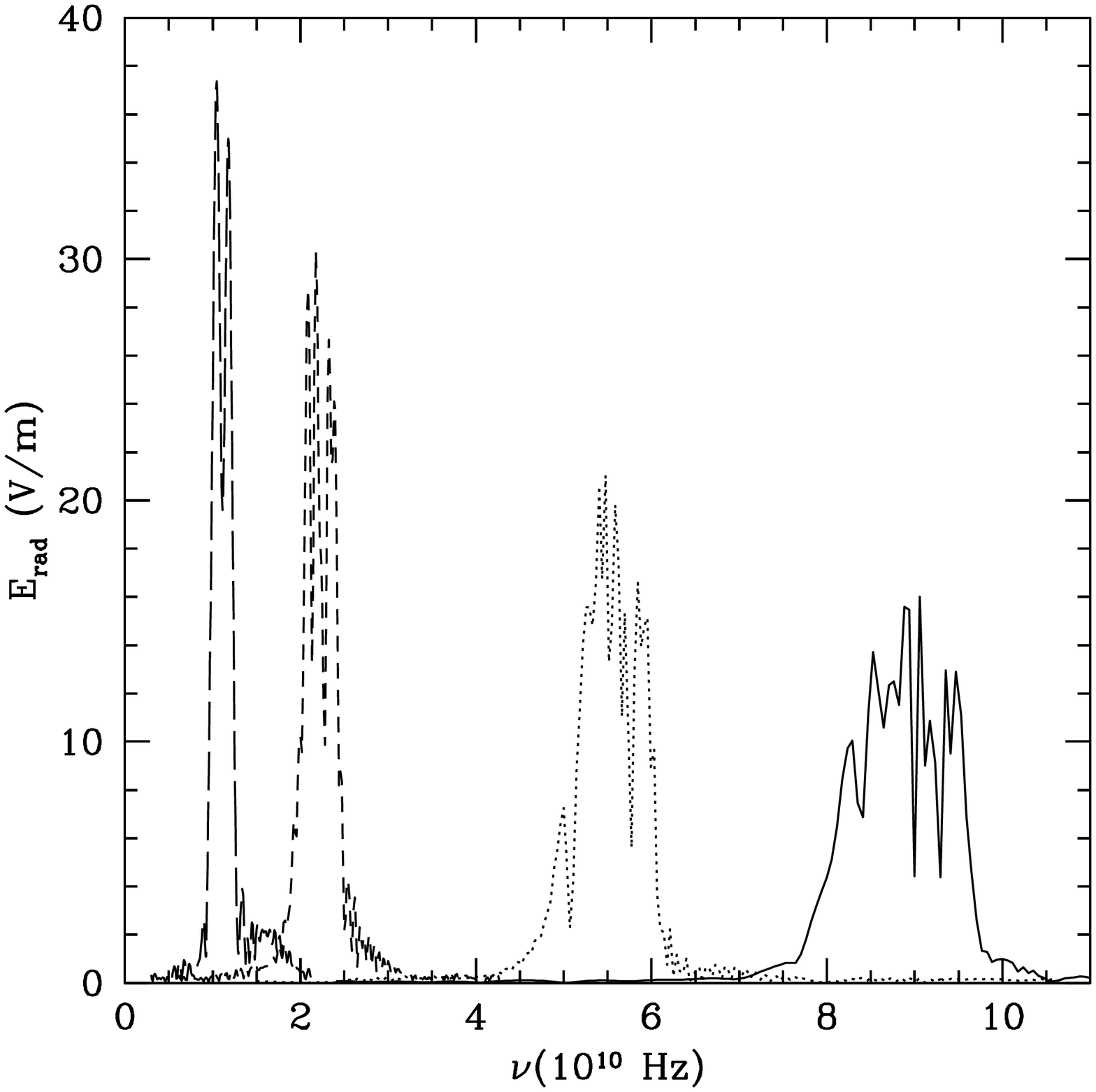}
    \mycaption a{Run~4: Spectrum for $\beta_{\textrm{w}}$ = 0.2 (solid), 0.5 (dotted), 0.8 (short-dashed), 0.9 (long-dashed). } 
    \label{fig:N200vfac_spec}
    \includegraphics[width=0.45\textwidth]{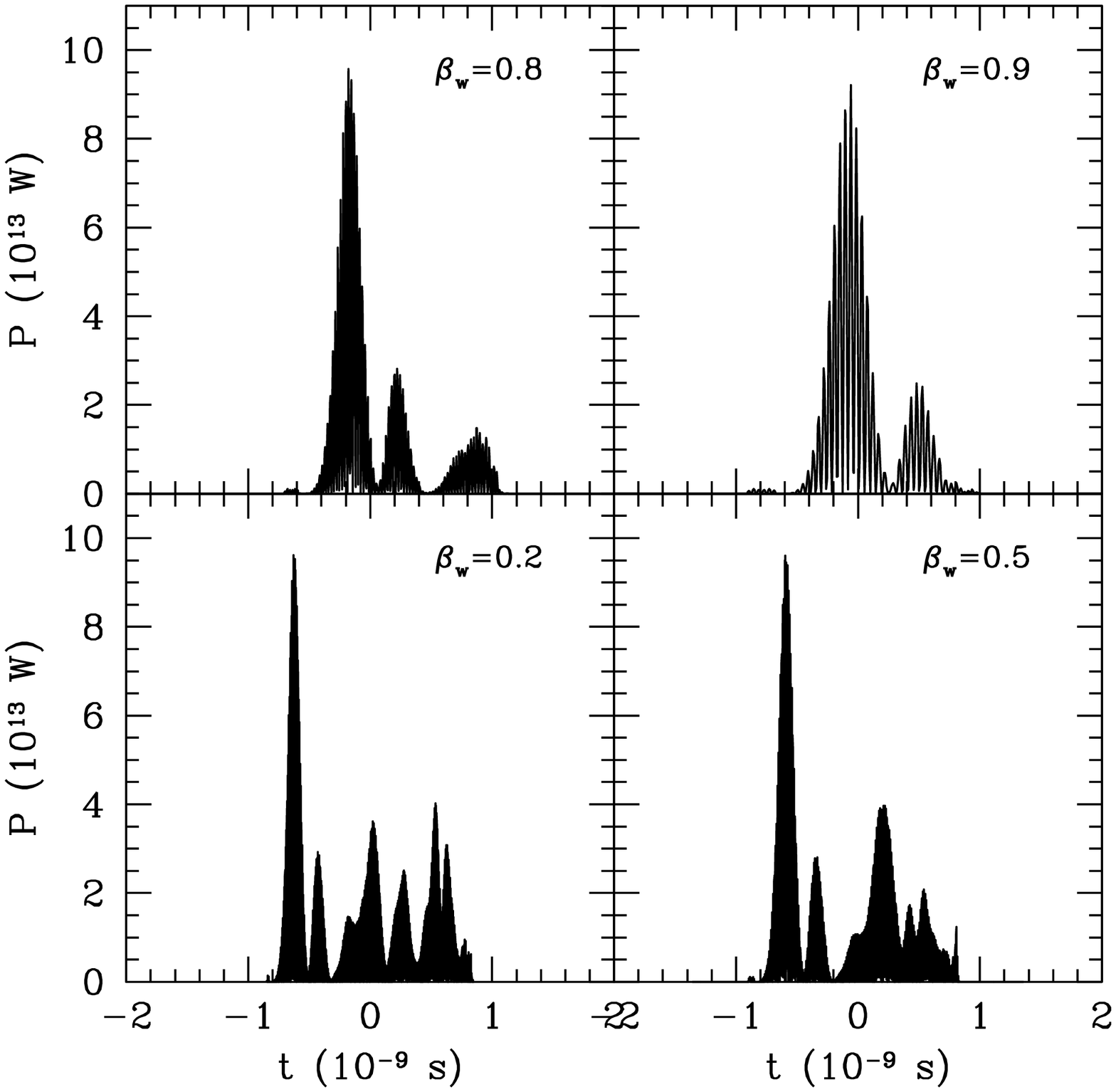}
    \mycaption b{Pulse power for Run~4 plotted for different
    $\beta_{\textrm{w}}$.} 
    \label{fig:N200vfac_power}
    \end{figure}

The radiation frequency in our simulations agrees well with the
expected resonance condition (compare eq.~(\ref{eq:nu_obs}) and Table~{\ref{tabel:inout}}). Therefore, by construction, the frequency of the emission is in the radio regime, between $1 \sim 10 \:\mbox{GHz}$. 
In some cases, the central frequency is lower than the resonance condition (Run~2-4: Fig.~\ref{fig:N200K_spec},~{\ref{fig:N200lamu_spec}},~{\ref{fig:N200vfac_spec}}). The shift in these cases is caused by multiple bunching where the average Lorentz factor decreases after each bunching, and therefore, also the frequency as determined by the resonance condition. 
As a result of multiple bunchings, the total bandwidth increases (Runs 3, 4, 6 in Figs.~{\ref{fig:N200lamu_spec}},~{\ref{fig:N200vfac_spec}},~{\ref{fig:n200nfac4_spec}}), whereas the relative bandwidth $\Delta \nu/\nu$ decreases. 

\subsubsection{Pulse shape, power, brightness temperature and radiation cone}
\begin{figure}[t!]
    \includegraphics[width=0.45\textwidth]{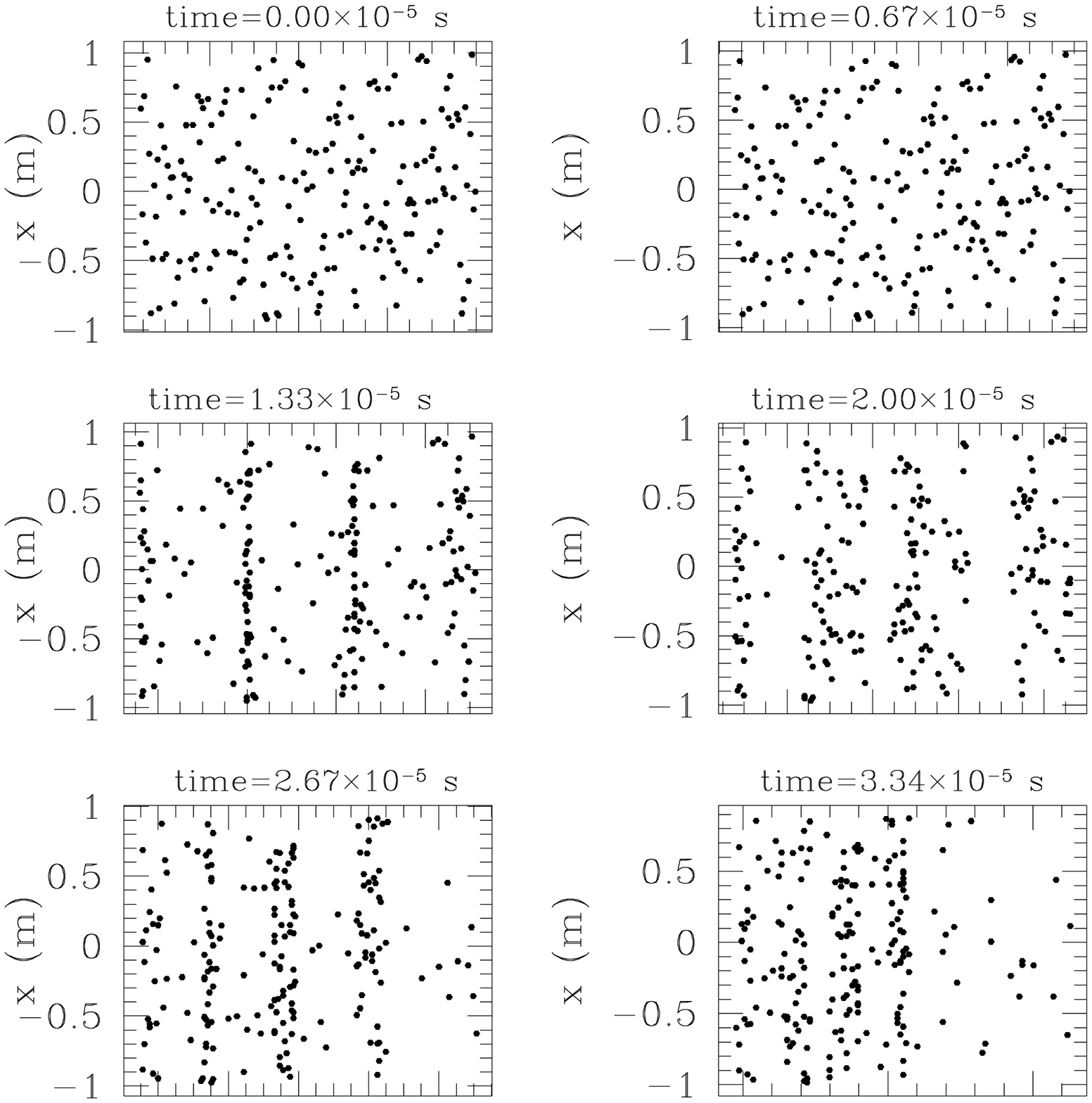}
    \mycaption a{Run~1: ($z$,$x$)-behaviour of the particles during their passage in the cavity (from left to rigth; from top to bottom). On the horizontal axis is plotted the $z$-coordinate of all particles, subsequently for z = 2, 4, 6, 8 and 10 km. Except for $t = 3.34 \cdot 10^{-5}\:\mbox{s}$, the horizontal spacing between the long ticks is 0.02 m (e.g. $\Delta z \approx 0.08 \:\mbox{m}$ for $t = 0\:\mbox{s}$). In the last plot, the spacing between the long ticks is 0.05 m. Bunching clearly occurs at $t = 1.33\cdot 10^{-5}\;\mbox{s}$, `de-bunching' at $t = 2.0 \cdot 10^{-5}\:\mbox{s}$ and `re-bunching' at $t = 2.67 \cdot 10^{-5}\:\mbox{s}$. }  
    \label{fig:bunching}
    \includegraphics[width=0.45\textwidth]{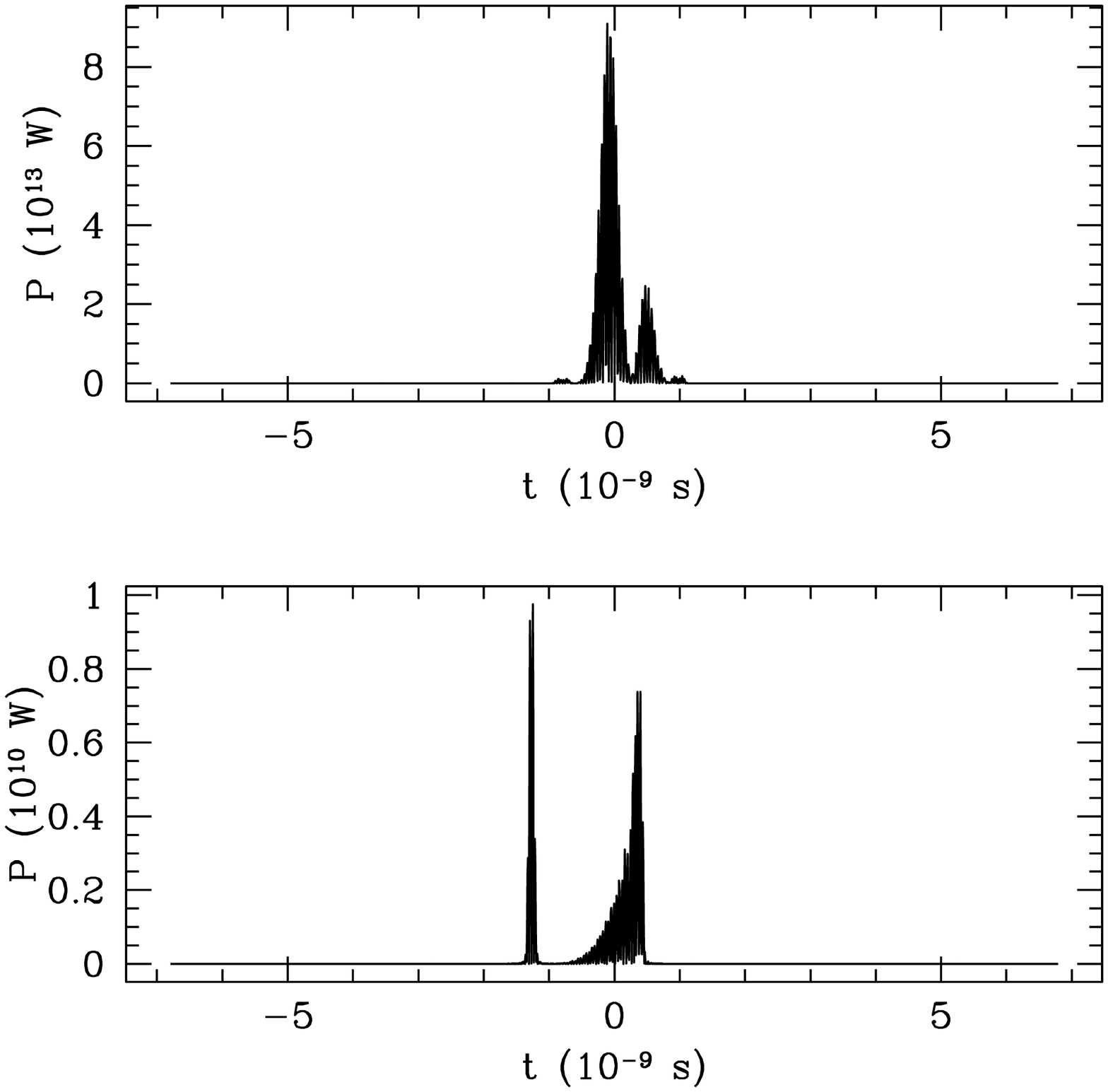}
    \mycaption b{Run~1: Power of the pulse at the end of the
    simulation $t = 3.3 \cdot 10^{-5}\:\mbox{s}$ (top: number of
    macroparticles N=200; bottom:
    N=2). On the horizontal axis is also plotted $t$, this is {\it not} the simulation time, but the {\it duration} of the pulse as a distant observer would measure. Negative $t$ corresponds to earlier arrival. 
The maximum power for $N=2$ (bottom) is $P_{2,\textrm{max}} = 9.75 \cdot 10^{9}\:\mbox{W}$. Since $P_{200} \propto N^2 P_{2}$, this implies the radiation pulse at the top is coherent. } 
    \label{fig:power}
    \end{figure}

\begin{figure}[t!]
    \includegraphics[width=0.45\textwidth]{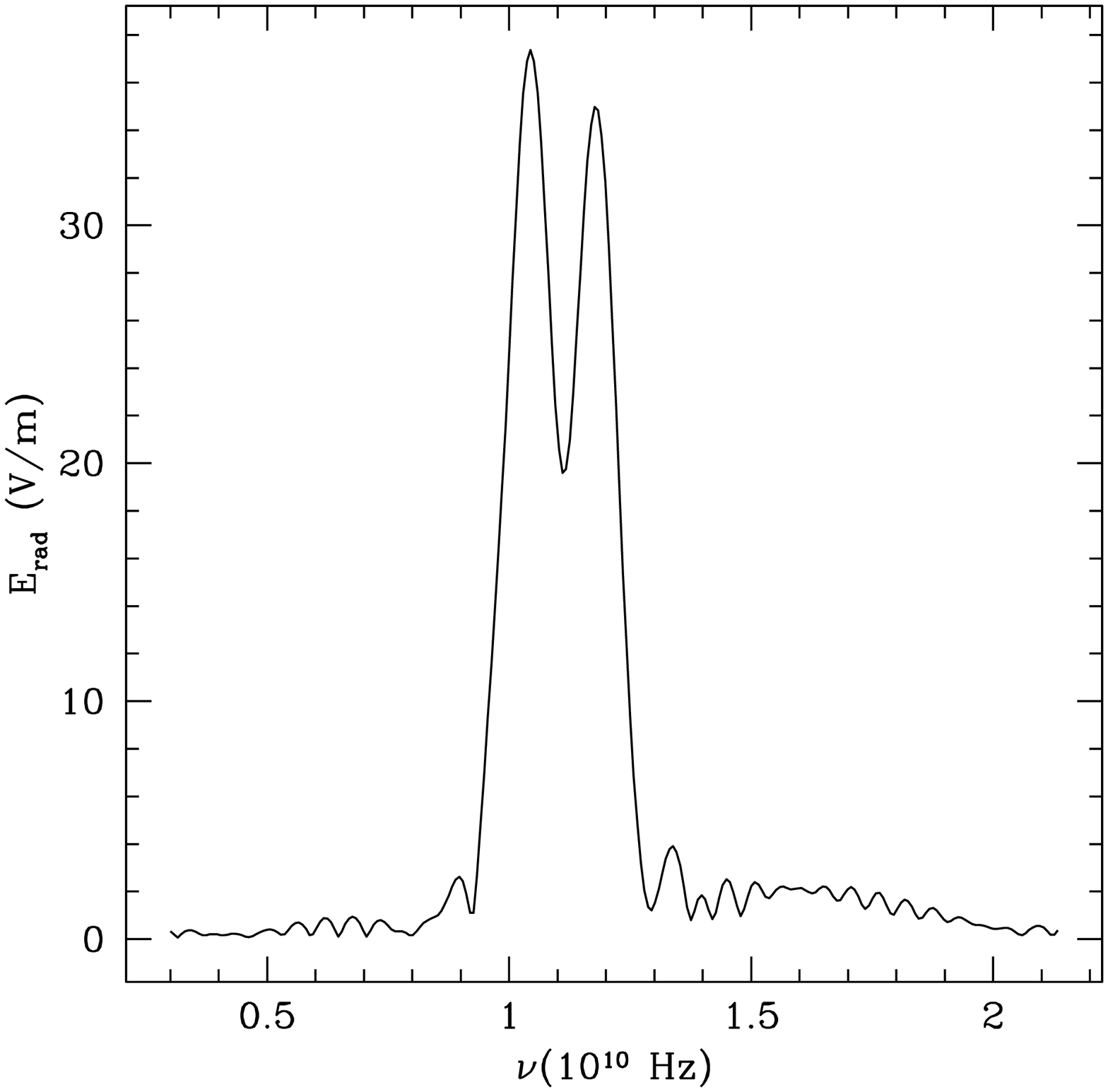}
    \mycaption c{Run~1: Spectrum at $t = 3.3 \cdot 10^{-5}\:\mbox{s}$. The
    central frequency is at $\nu = 11\:\mbox{GHz}$, with FWHM of $\Delta \nu
    = 2.42\:\mbox{GHz}$. The peak on the right, centered on $\nu_{\textrm{res}} = 11.9\:\mbox{GHz}$, is caused by the first bunching at $t = 1.33 \cdot 10^{-5}\:\mbox{s}$. And the peak on the left corresponds to the second buching, where $\gamma$ dropped $5\%$, resulting in a lower $\nu_{\textrm{res}}$. }  
    \label{fig:spectrum}
    \includegraphics[width=0.45\textwidth]{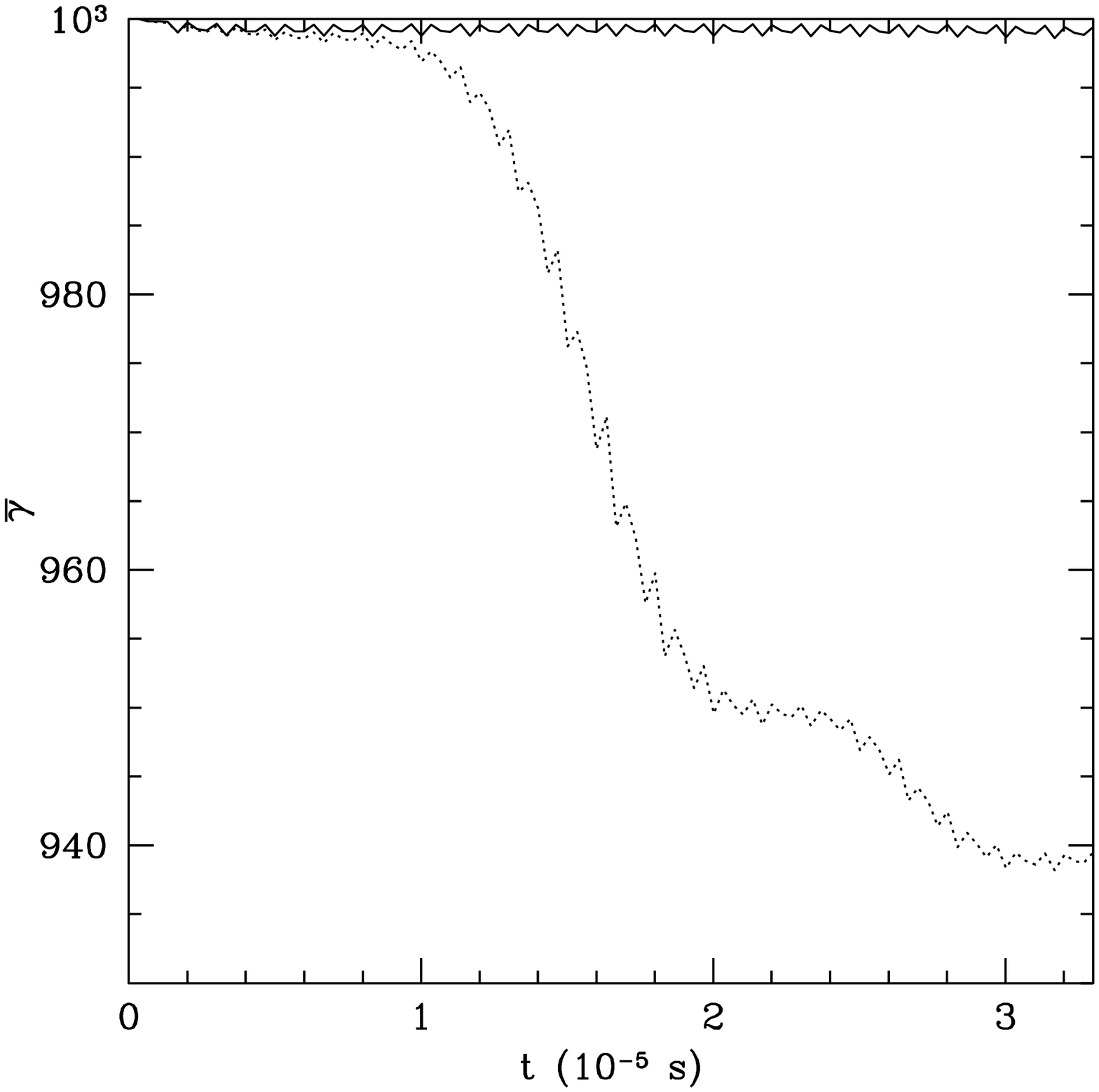}
    \mycaption d{Run~1: Time development of average Lorentz
    factor (i.e. per
    macro-particle) in two simulations, one with $N=2$ (top) and the other
    with $N=200$ (bottom). The $N=2$ case reflects the
    behaviour of particles radiating incoherently, with constant energy
    loss over time as compared with the bunching case, where
    there is a steep drop in energy ($\Delta \bar{\gamma} = 50$) of the beam particles between 1.0 and 1.8 $\cdot 10^{-3}\:\mbox{s}$. Between $2.4 \cdot 10^{-5}\:\mbox{s}$ and $2.8 \cdot 10^{-5}\:\mbox{s}$ another drop in beam energy occurs, which corresponds to a second bunching of the particles. This time, the beam energy loss is less than for the first time, and the duration is much shorter. The pulse power (Fig.~{\ref{fig:power}}) reflects these characterics where the first pulse is more powerful and has a longer duration than the second pulse. }  
    \label{fig:avggamma}
    \end{figure}

\begin{figure}[t!]
    \includegraphics[width=0.45\textwidth]{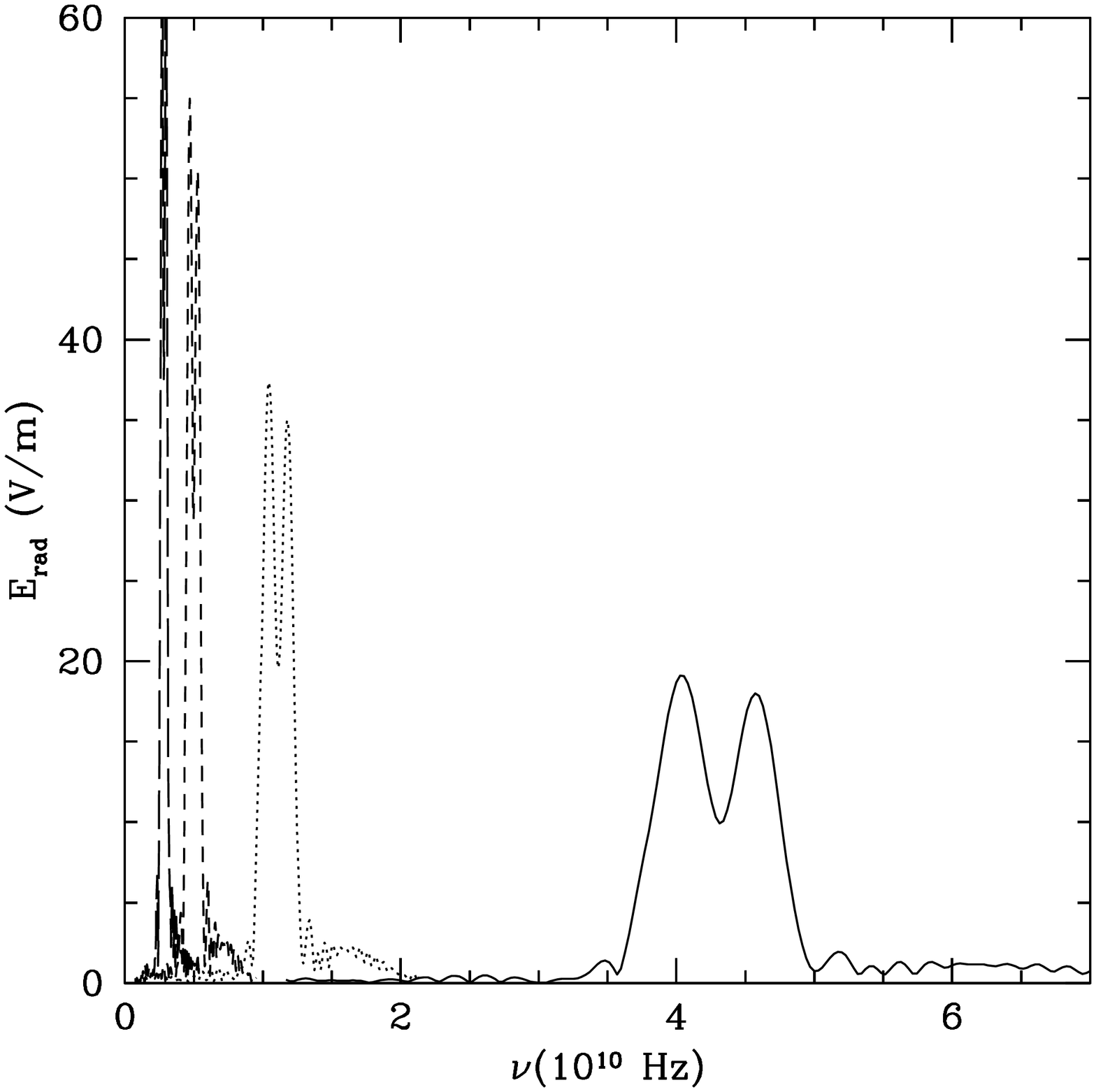}
    \mycaption a{Spectra for Run~2 plotted over
      each other; we have $K = 5$ (solid), 10 (dotted), 15 (short-dashed) and 20 (long-dashed). The resonance frequency shifts according to (\ref{eq:nu_obs}).The bandwidth increases with increasing $\nu_{\textrm{res}}$.}
    \label{fig:N200K_spec}
    \includegraphics[width=0.45\textwidth]{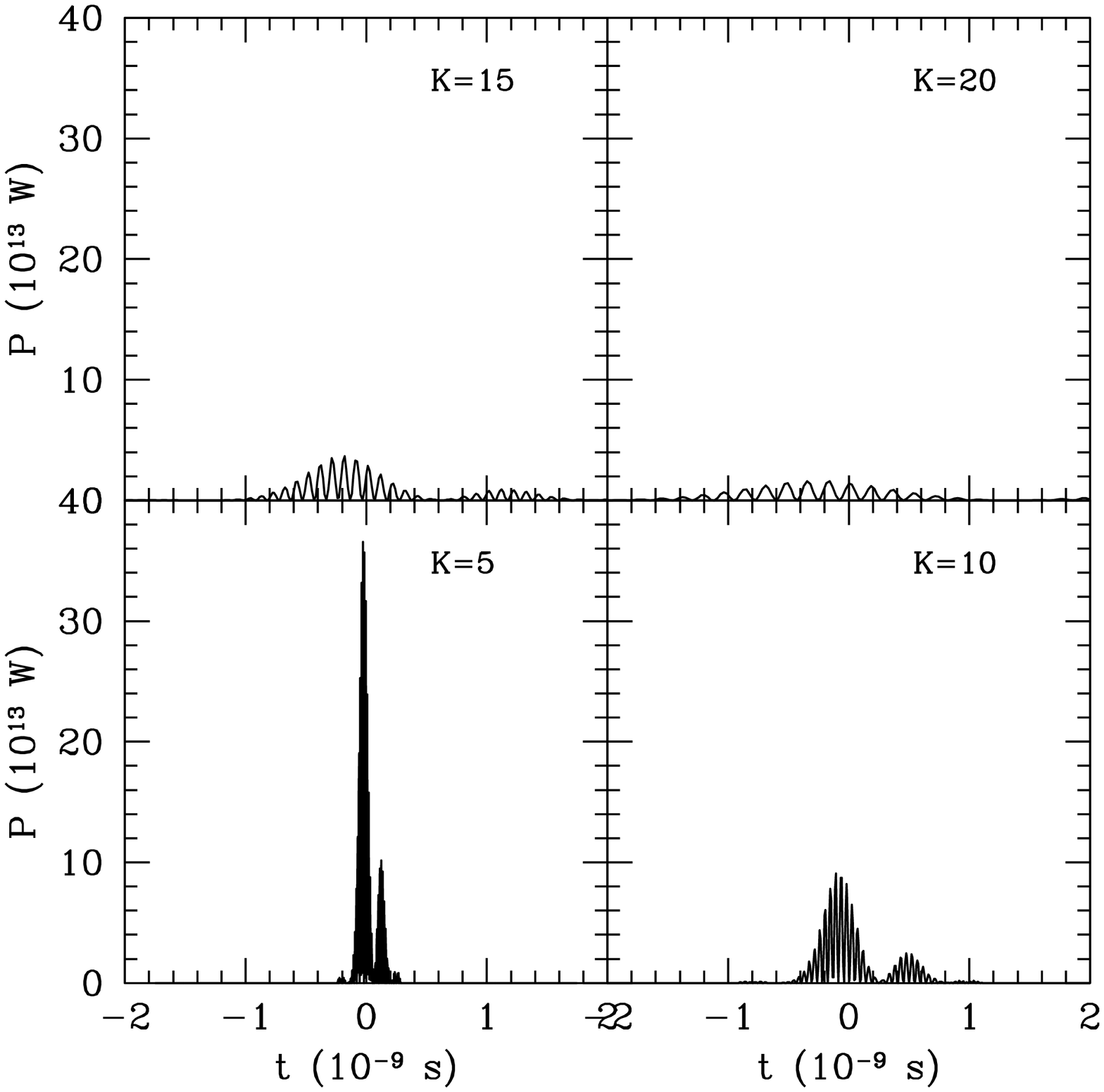}
    \mycaption b{Pulse power for Run~2 plotted for different $K$. The maximum scales roughly as $K^{-2}$. }
    \label{fig:N200K_power}
    \end{figure}

\begin{figure}[t!]
    \includegraphics[width=0.45\textwidth]{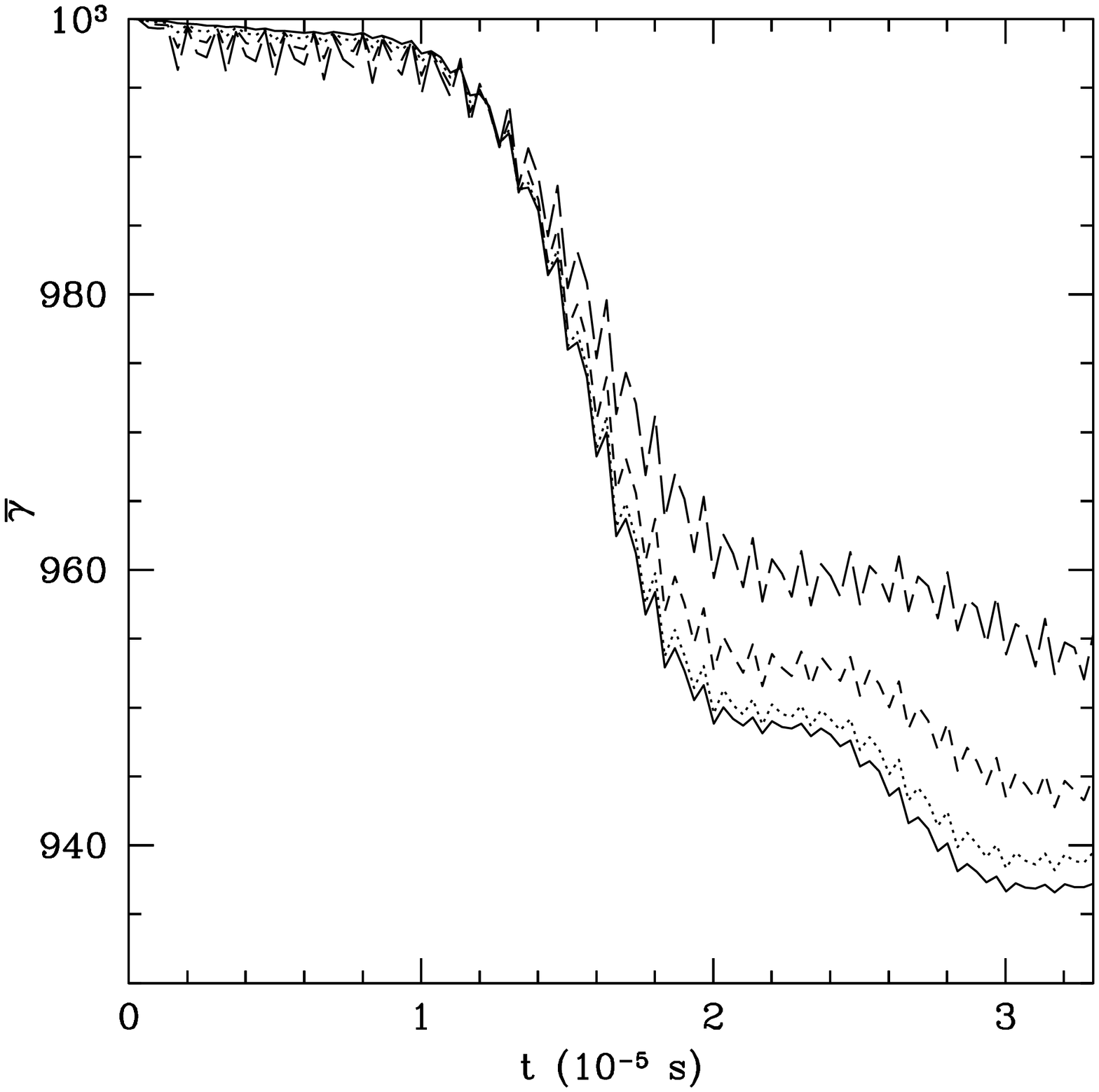}
    \mycaption c{Run~2: ($\bar{\gamma}$,$t$)-plot for $K = 5$, 10, 15, 20 (labelling as in Fig.~\ref{fig:N200K_spec}). Obviously, for all $K$, the beam particles bunch twice (also, see Fig.~{\ref{fig:N200K_power}}). corresponding
to the maximum power, the beam energy loss is largest here for $K = 5$. } 
    \label{fig:N200K_tgam}
    \end{figure}

\begin{figure}[t!]
    \includegraphics[width=0.45\textwidth]{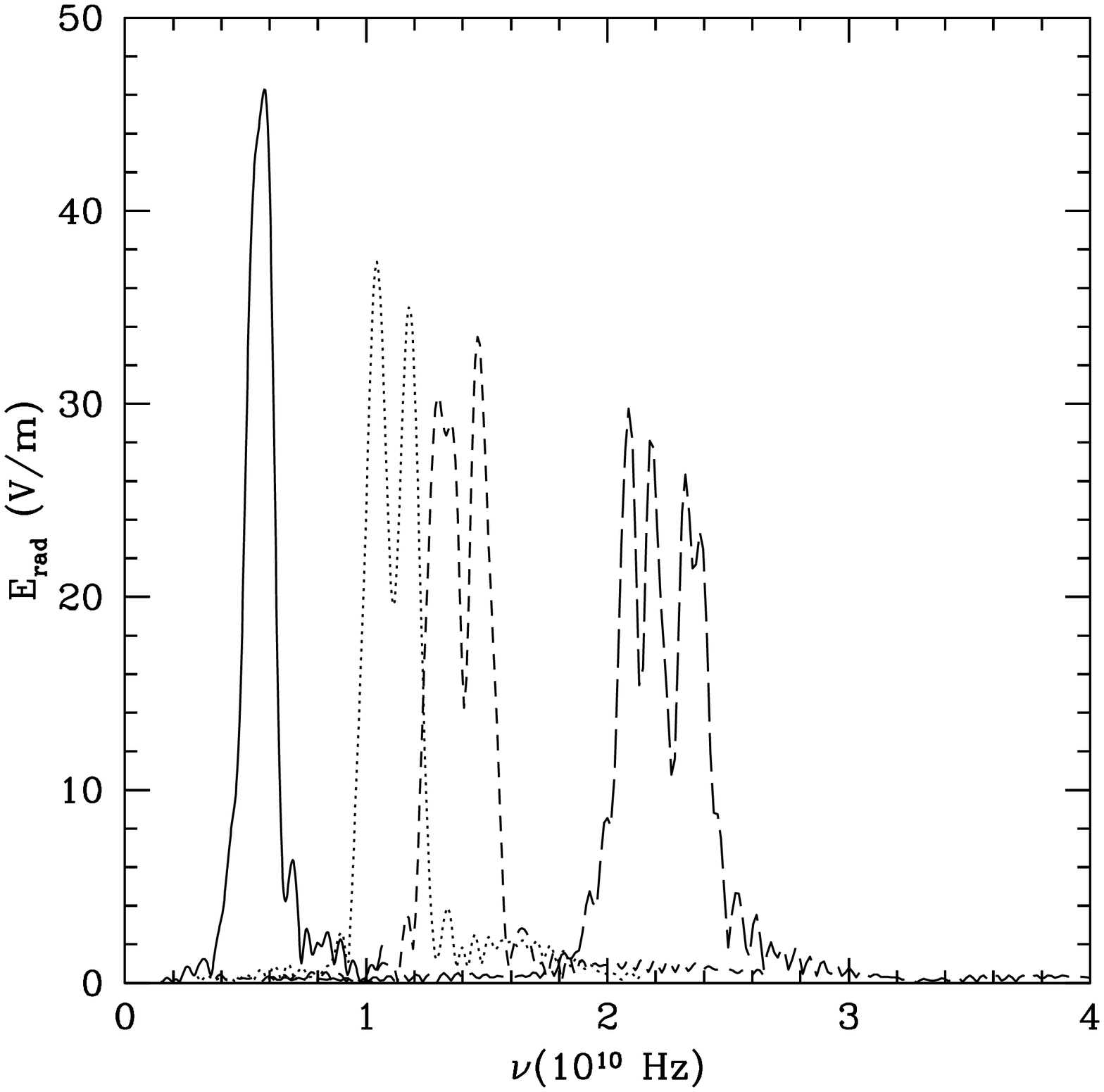} 
    \mycaption a{Run~3: Spectrum at the end of the simulation for
    different 
      $\lambda_{\textrm{w}} = 100$ m (solid), $50$ m (dotted), $40$ m (short-dashed), $25$ m (long-dashed). Spectral broadening is due to more bunchings occuring during the simulation. } 
    \label{fig:N200lamu_spec} 
    \includegraphics[width=0.45\textwidth]{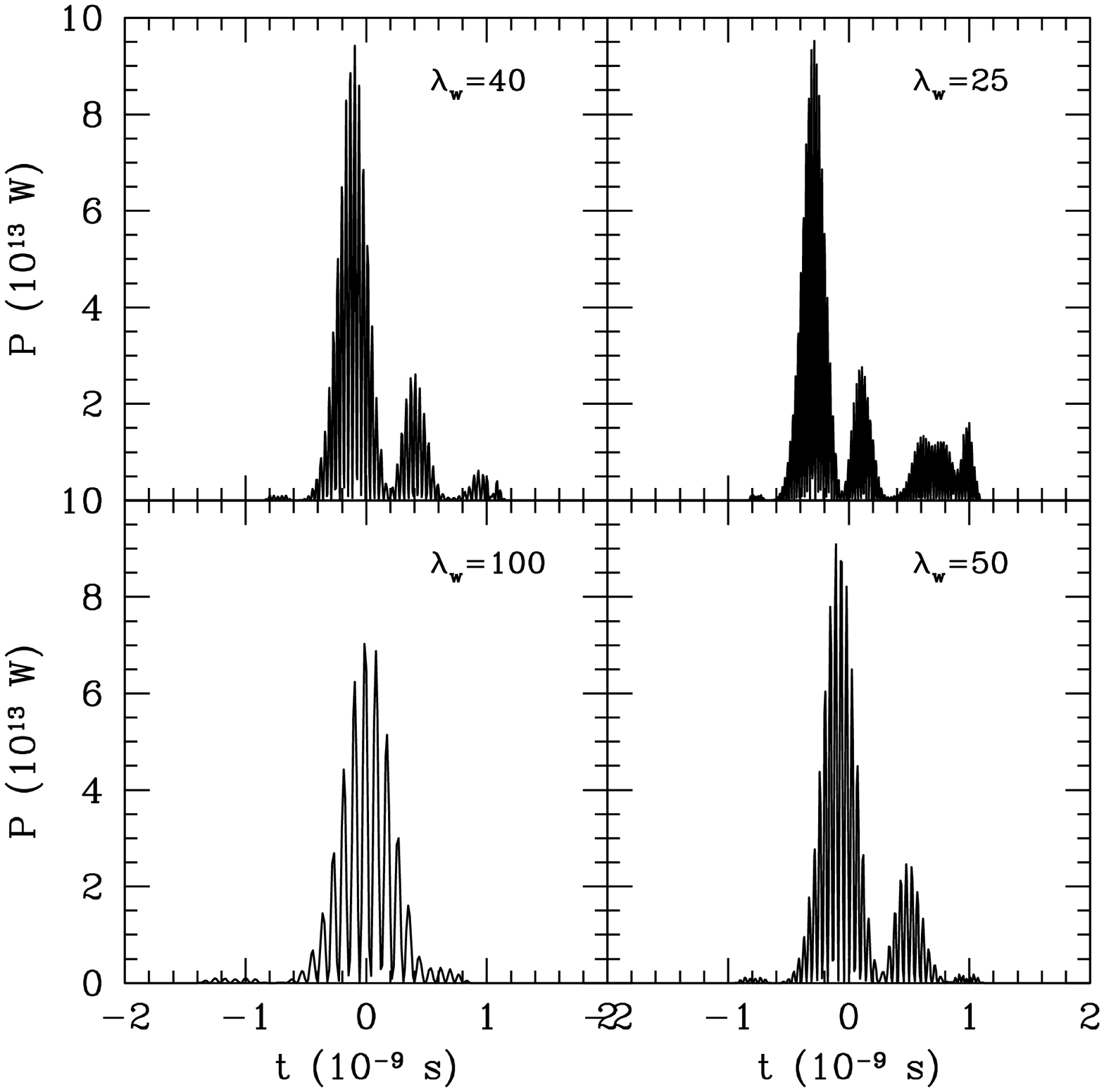} 
    \mycaption b{Pulse power for Run~3 plotted for different $\lambda_{\textrm{w}}$ values. Maximum powers are of about the same order of magnitude. Only the number of pulses differs for each run, due to number of bunching times.  }
    \label{fig:N200lamu_power} 
    \end{figure}

\begin{figure}[t!] 
    \includegraphics[width=0.45\textwidth]{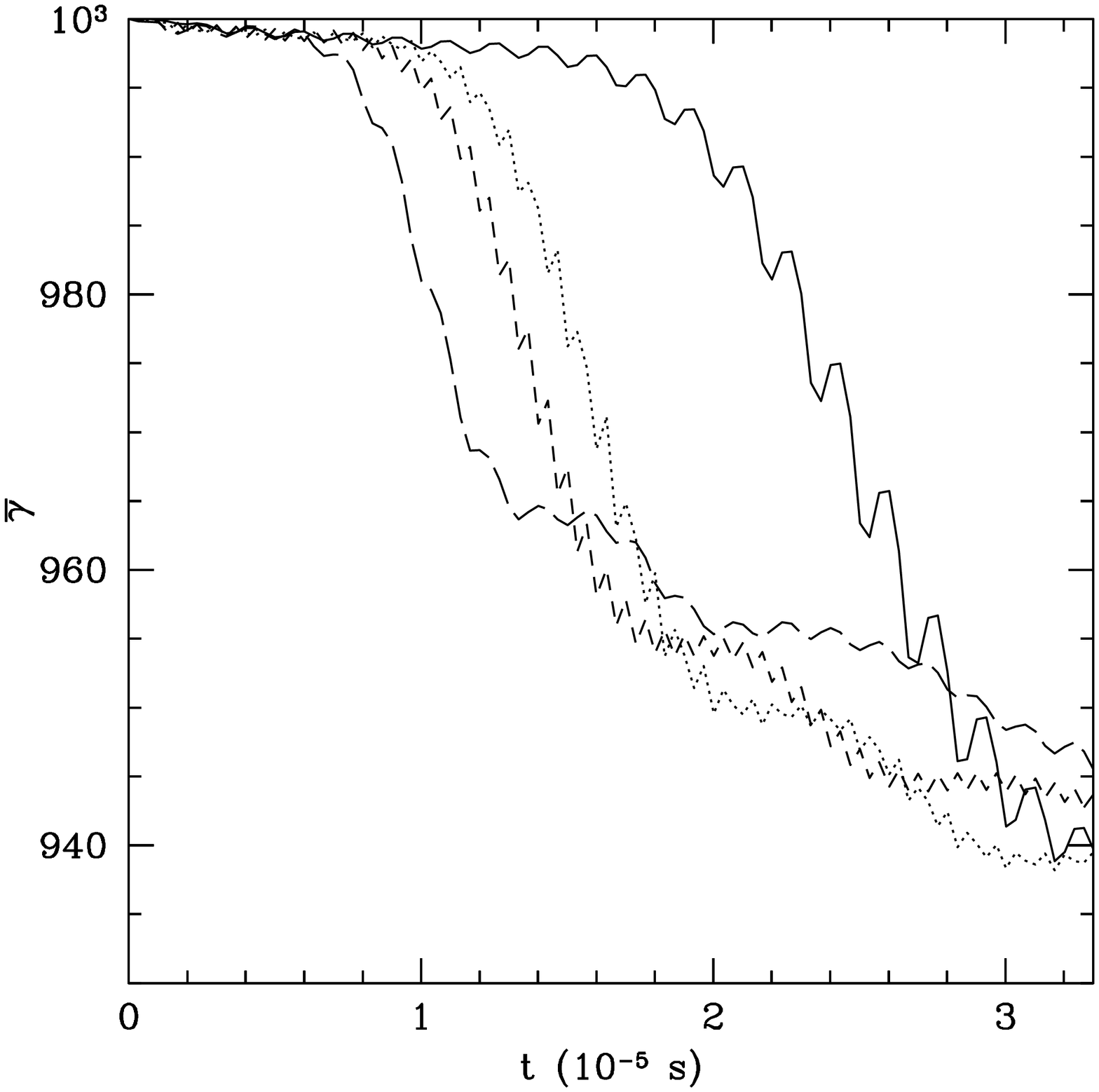} 
    \mycaption c{Run~3: The average Lorentz factor of the
    macroparticles during 
      their passage in the cavity for $\lambda_{\textrm{w}} = 100$, $50$, $40$, 
      $25$ m (labelling as in Fig.~{\ref{fig:N200lamu_spec}}). The starting time of the first bunching for each run is about $10 \:t_{\textrm{eff}}$ (\ref{eq:teff}). The energy beam loss after the first bunching is about $5\%$. } 
    \label{fig:N200lamu_tgam} 
    \end{figure}

\begin{figure}[t!]
    \includegraphics[width=0.45\textwidth]{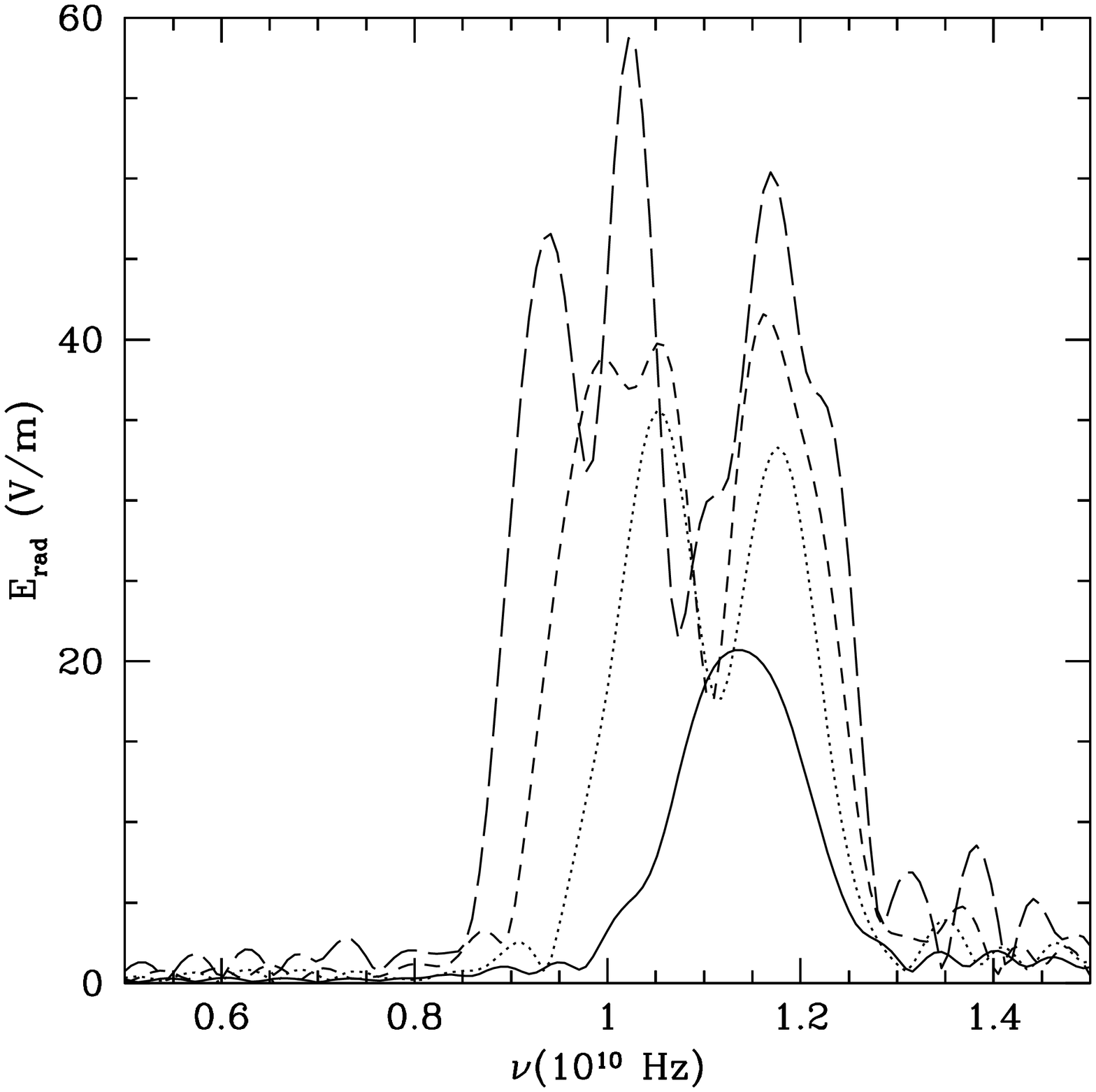}    
    \mycaption a{Run~6: Spectrum (from top to bottom) for $n/n_{\textrm{GJ}\ast}$ = 0.1 (solid), 0.2 (dotted), 0.3 (short-dashed) and 0.4 (long-dashed). The spectrum broadens as the number density increases, as expected from Fig.~{\ref{fig:n200nfac4_tgam}}, which shows single bunching for $n/n_{\textrm{GJ}} =
0.1$, but multiple
bunching for $n/n_{\textrm{GJ}} > 0.1$. }
    \label{fig:n200nfac4_spec}
    \includegraphics[width=0.45\textwidth]{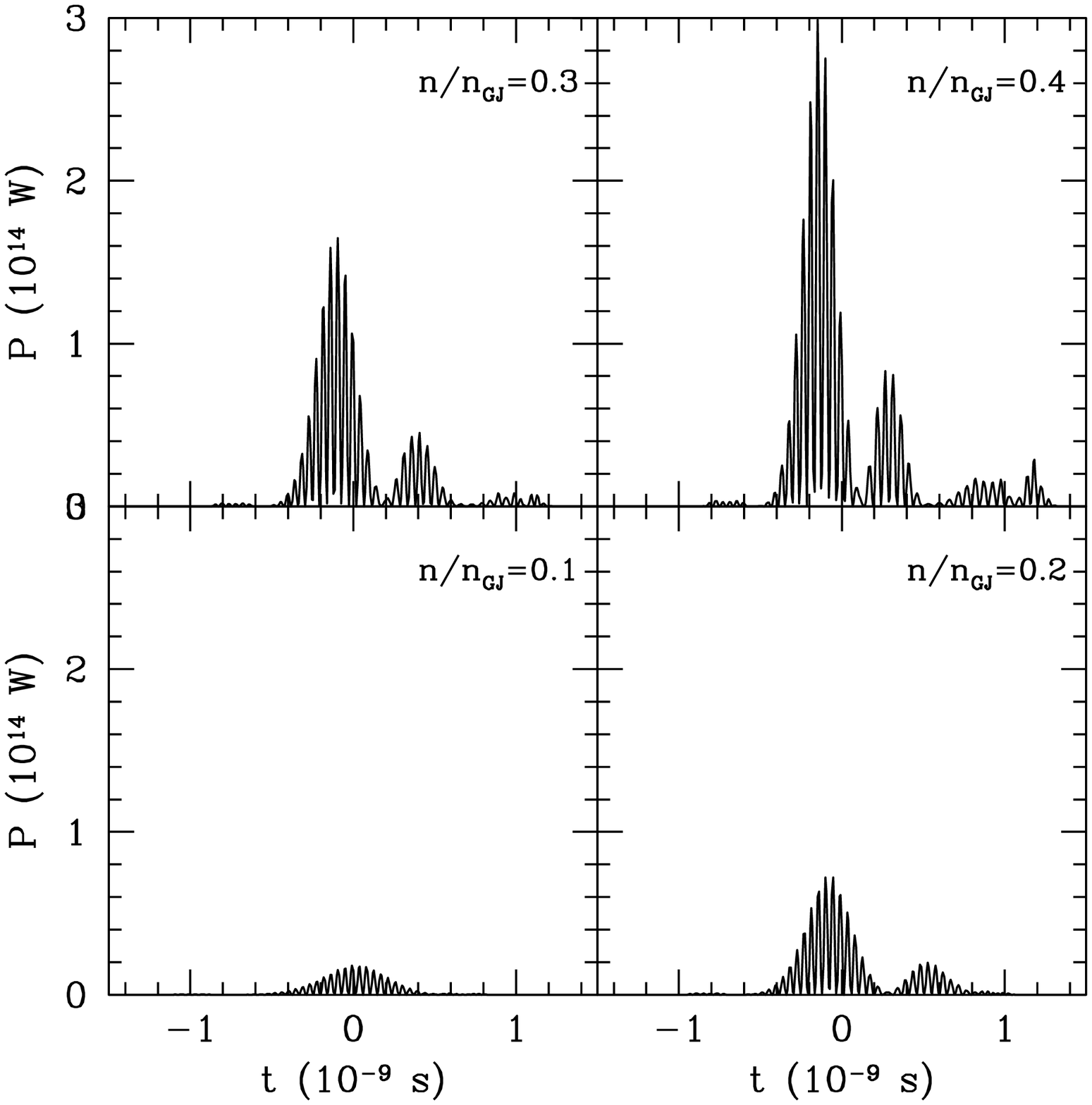}
    \mycaption b{Pulse power for Run~6 plotted for different number densities relative to $n_{\textrm{GJ}\ast}$. As expected, the maximum power scales as $n^2$. }
    \label{fig:n200nfac4_power}
    \end{figure}

\begin{figure}[t!]
    \includegraphics[width=0.45\textwidth]{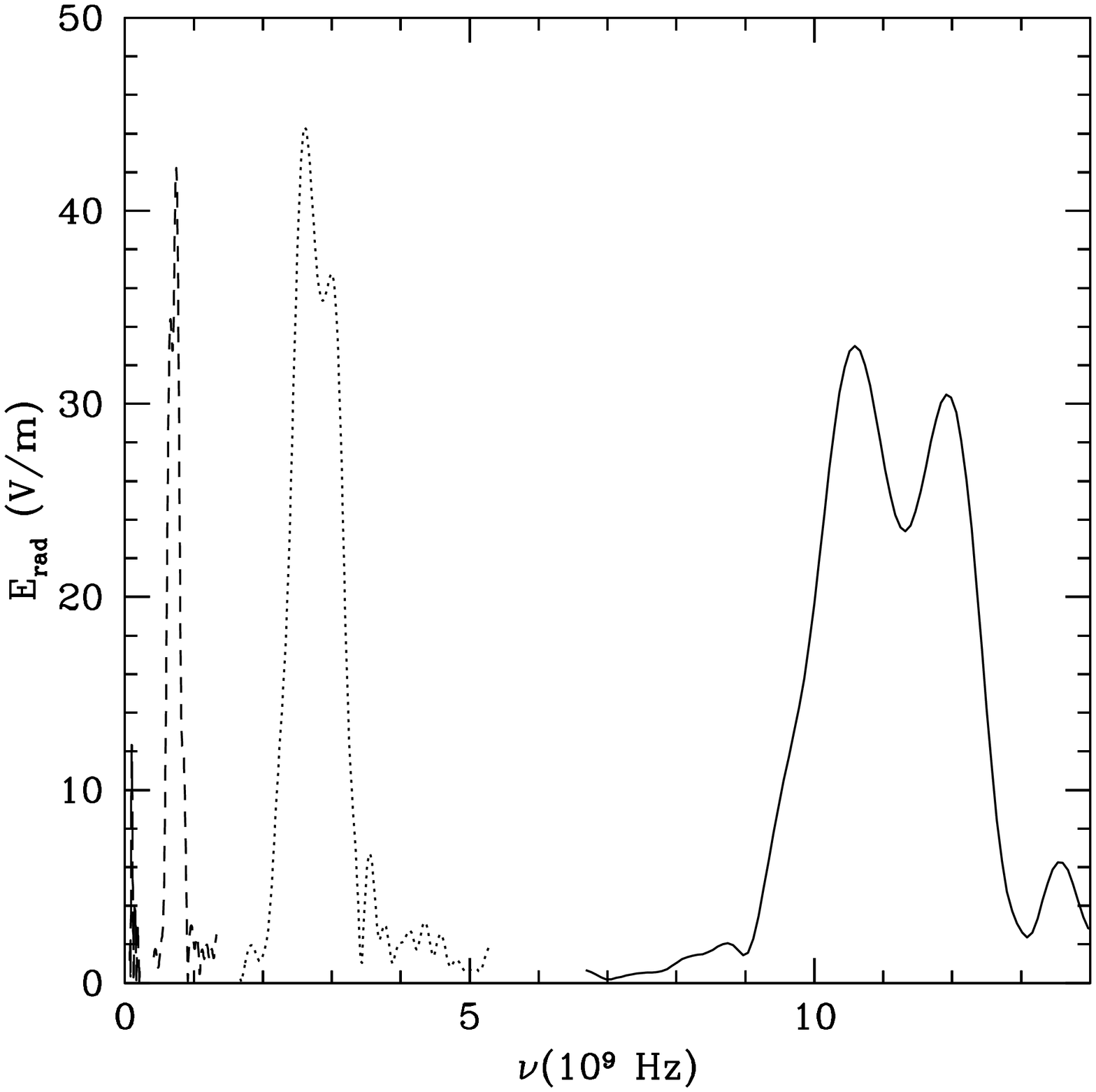}
    \mycaption a{Run~7: Spectrum for initial Lorentz factor
of the beam particles $\gamma = 1000$ (solid), 500 (dotted0), 250 (short-dashed), 100 (long-dashed). The central frequency shift as $\gamma^2$. }
    \label{fig:n200gamma_spec}
    \includegraphics[width=0.45\textwidth]{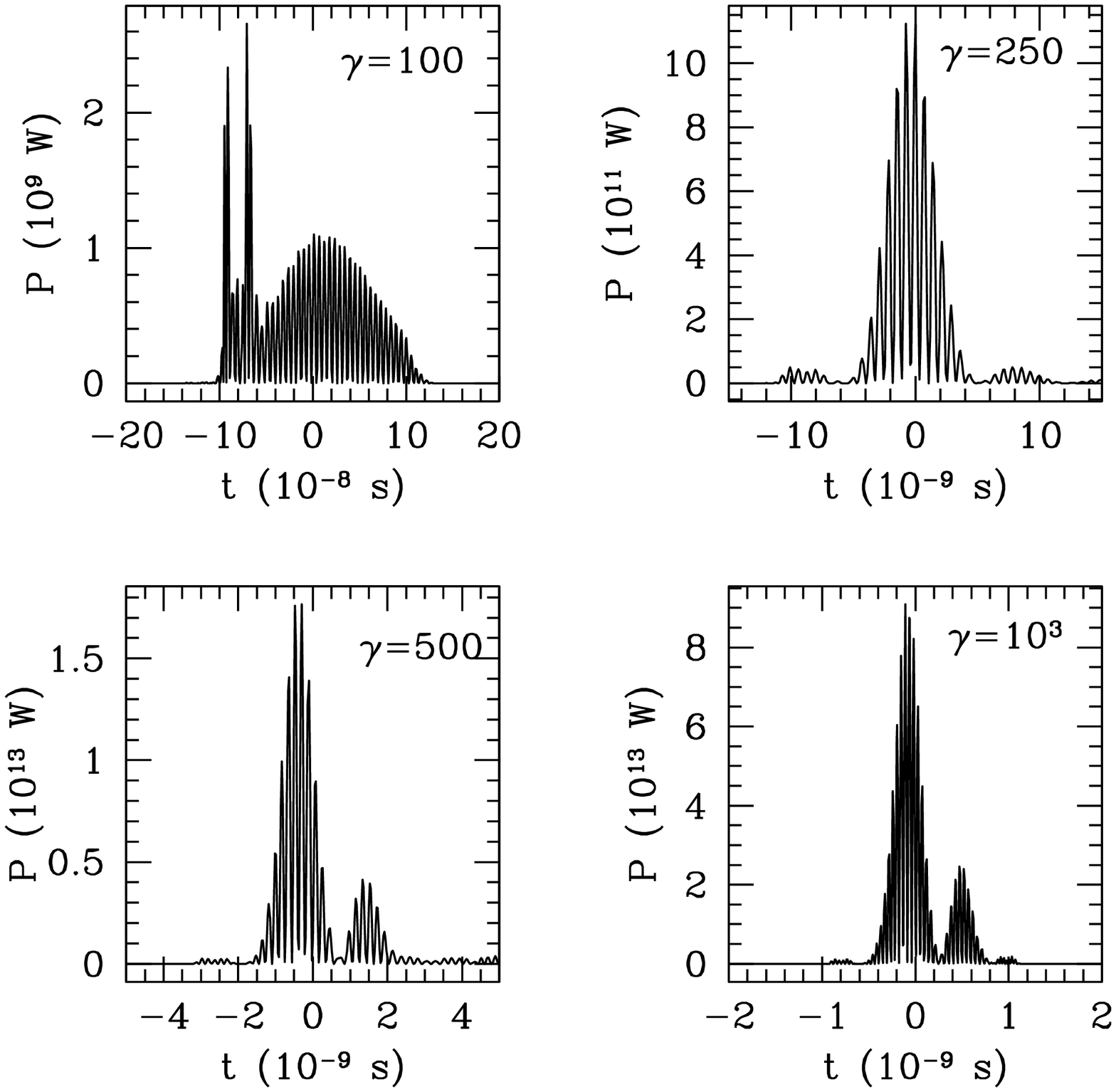}
    \mycaption b{Pulse power for Run~7 plotted. }
    \label{fig:n200gamma_power}
    \end{figure}

The pulse shapes of the coherent radiation can be found in
Figs.~{\ref{fig:power}}, {\ref{fig:N200K_power}}, {\ref{fig:N200lamu_power}, {\ref{fig:N200vfac_power}}, {\ref{fig:n200nfac4_power}} and {\ref{fig:n200gamma_power}}}. \\ The
number of temporal peaks corresponds to the number of bunching times (not the
number of bunches that is formed); e.g. in Run~1 ($N=200$) the beam
particles bunched twice (Fig.~{\ref{fig:avggamma}}),
and the pulse has two peaks (Fig.~{\ref{fig:power}}). The maximum pulse power always occurs at the first time bunching. The following bunch formations result in less powerful pulses (though not completely incoherent). The maxima of the pulses have the same order of magnitude for most runs (Table~{\ref{tabel:inout}}), except for $\gamma = 100$, where the coherent emission becomes less
efficient (and also for $B_0 = 0.05\:\mbox{T}$, but this is discussed in the next section). The optimum efficiency of the emission power is reached for
$\gamma \simeq 300$ as follows from the ratio $\epsilon_{\textrm{p}}/\epsilon_{\textrm{b}}$ (Table~{\ref{tabel:inout}}, Run~7: Fig.~{\ref{fig:n200gamma_power}})  \\
The structure within {\it each} pulse reflects the wiggling motion of
the particles (e.g. Fig.~{\ref{fig:N200lamu_power}}, where $\lambda_{\textrm{w}} = 100\:\mbox{m}$). \\
The typical duration of the pulse is a few nanoseconds.\\ 
The brightness temperature $T_{\textrm{b}}$ is derived using 
    \begin{equation}
    I = \frac{k_{\textrm{B}}T_{\textrm{b}}\nu^2}{c^2} = \frac{W/\Delta
    t}{A \Delta\nu \Delta\Omega}
    \end{equation}
where $I$ is the radiation intensity, $k_{\textrm{B}}$ is Boltzmann's
constant, $W/\Delta t$ is the power, which we take as
$P_{\textrm{max}}$, $A$ is the emitting surface, which corresponds to
the beam cross-section, $\Delta \nu$ is the bandwidth, and $\Delta
\Omega = \pi \tan^2\phi$ the solid angle into which the radiaton is
emitted (\ref{eq:openangle}). \\
Converting the maximum power into the brightness temperature, we find
that, again, for most cases the brightness temperature is approximately $10^{30}\:K$ (Table~{\ref{tabel:inout}}). 

\subsection{Background magnetic field}
\begin{figure}[t!]
    \includegraphics[width=0.45\textwidth]{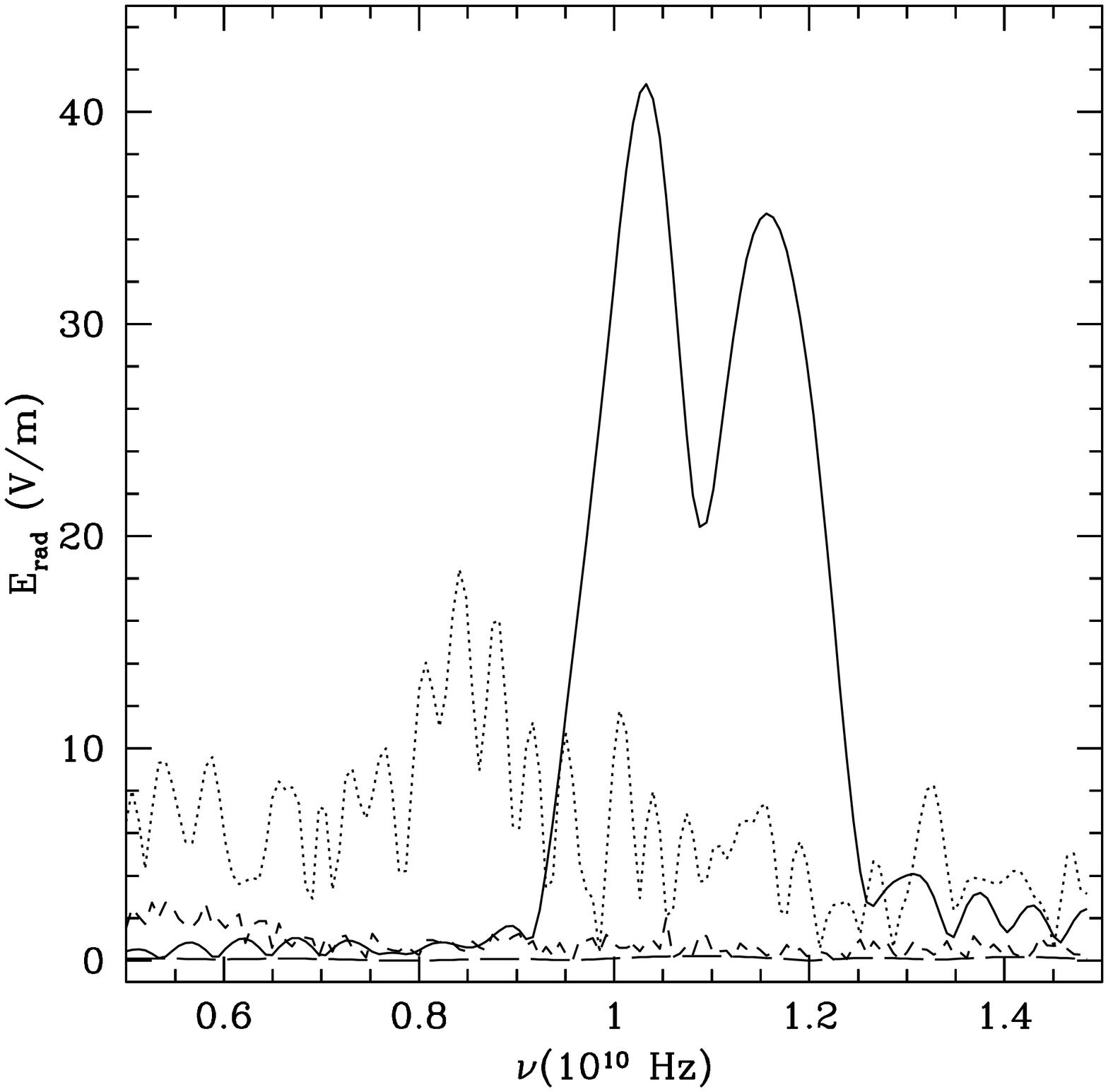}
    \mycaption a{Run~5: Spectra for (from top to bottom along the left vertical axis) $B_0 = 10^{-3}$ T (solid), $10^{-2}$ T (dotted), $0.0025$ T (short-dashed), $0.050$ T (long-dashed). For $B_0 = 10^{-3}\:\mbox{T}$, the spectrum shows the same properties (central frequency and bandwidth) as in the case where the background magnetic field is absent. For $B_0 = 10^{-2}\:\mbox{T}$, the central frequency shifts to $2\:\mbox{GHz}$, whereas the bandwidth drops to $0.61\:\mbox{GHz}$.}
    \label{fig:N200Ball_spec}
    \includegraphics[width=0.45\textwidth]{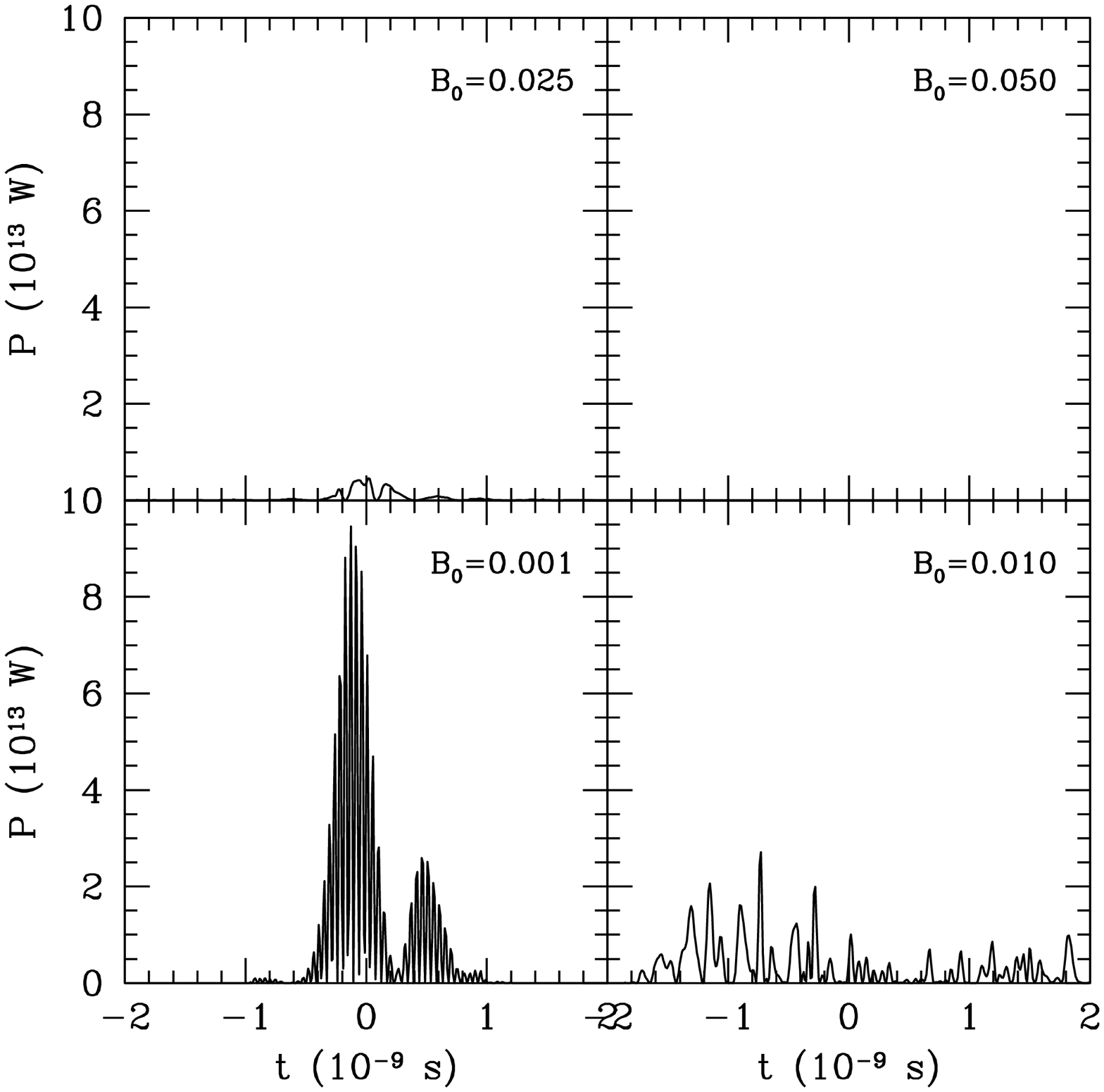}
    \mycaption b{Pulse power (logarithmic) for Run~5. The power of the radiation becomes smaller for $B_0 = 0.025$ and 0.05 T. }
    \label{fig:N200Ball_power}
    \end{figure}

\begin{figure}[t!]
    \includegraphics[width=0.45\textwidth]{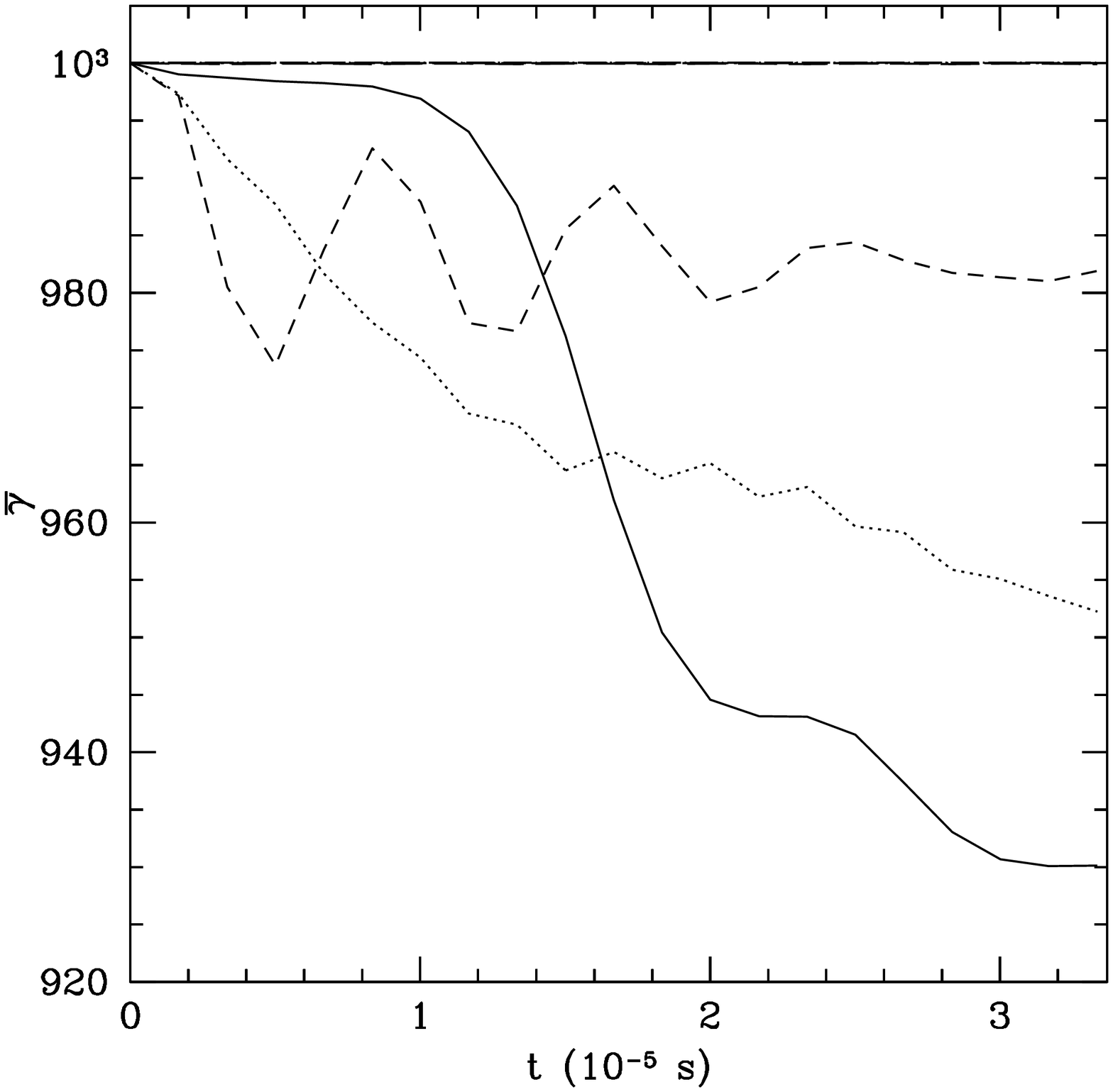}
    \mycaption c{Run~5: The average Lorentz factor versus the average
    distance $z$ of the
      particles in simulations with background magnetic field
    for $B_0 = 10^{-3}, 10^{-2}, 2.5 \cdot 10^{-2}, 5.0 \cdot 10^{-2}\:\mbox{T}$ (labelling as in Fig.~{\ref{fig:N200Ball_spec}}). The last case
coincides practically with the horizontal line $\bar{\gamma} = 1000$. 
Only for $B_0 = 10^{-3}\:\mbox{T}$ and $B_0 = 10^{-2}\:\mbox{T}$ the beam particles show FEL action. } 
    \label{fig:N200Ball_tgam}
    \end{figure}

An ambient magnetic field in the $z$-direction (Run~5) can have a drastic effect on the bunching (Fig.~{\ref{fig:N200Ball_tgam}}). 
For a background field just smaller than the wiggler field, $B_0 = 10^{-3}\:\mbox{T}$, the particles' behaviour shows no difference with the simulations where the background field is absent. For $B_0 = 10^{-2}\:\mbox{T}$, we still observe bunching. For larger values of $B_0$, however, bunching disappears completely. Actually, such behaviour is expected from equation~(\ref{eq:b0maximum}): since the motion of the particles is no longer free in the transverse direction
, the ponderomotive force is less effective and the resulting bunching is less
pronounced. For all runs, the angle between the particles' velocity and the total magnetic field ${\bf B} = B_0 + B_{\textrm{w}}$, which is proportional
to ${\bf v} \cdot {\bf B}$ is indeed found to be nonzero. \\
Since only the cases where $B_0 \leq 10^{-2}\:\mbox{T}$ show bunching, these are discussed further. 
The spectrum for this run is plotted in Fig.~{\ref{fig:N200Ball_spec}}. Only for $B_0 = 10^{-3}\:\mbox{T}$, the characteristics of the spectrum are similar to the cases where $B_0 = 0 \:\mbox{T}$. For $B_0 = 10^{-2}\:\mbox{T}$, the resonance frequency shifts to $\nu_{res} = 2.18 \:\mbox{GHz}$ (as compared to the expected value of $11.9\:\mbox{GHz}$). The bandwidth is $0.61\:\mbox{GHz}$, and is much smaller than before.

    \begin{figure}[t!]
    \includegraphics[width=0.45\textwidth]{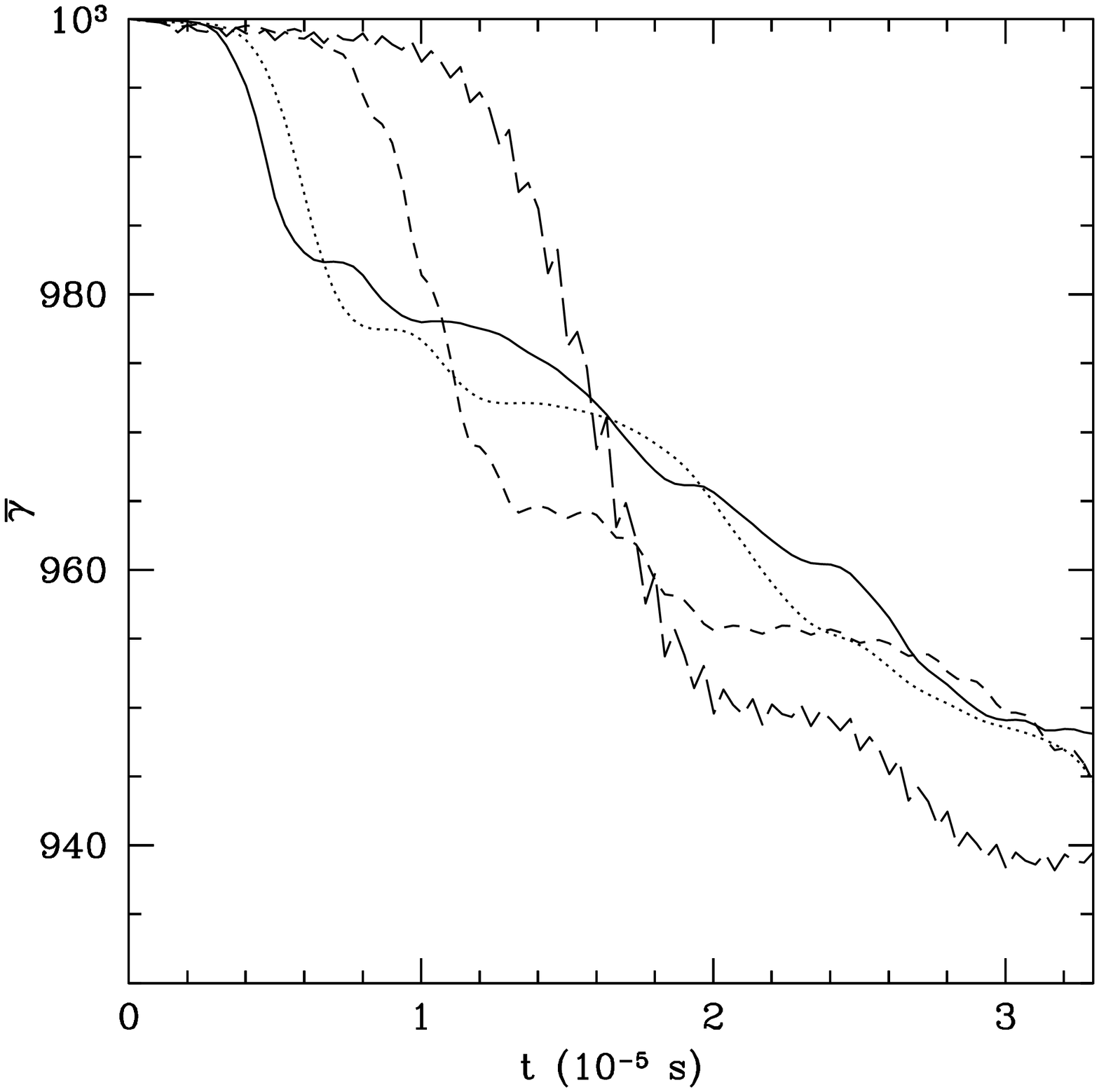}    
    \mycaption c{Run~4: The average Lorentz factor versus the average
    distance $z$ of the
      particles in simulations with changing $\beta_{\textrm{w}} = 0.2$ , 0.5, 0.8, 0.9 (labelling same as Fig.~{\ref{fig:N200vfac_spec}}). The starting time of the first bunching is $10\:t_{\textrm{eff}}$. }
    \label{fig:N200vfac_tgam}
    \end{figure}

    \begin{figure}[t!]
    \includegraphics[width=0.45\textwidth]{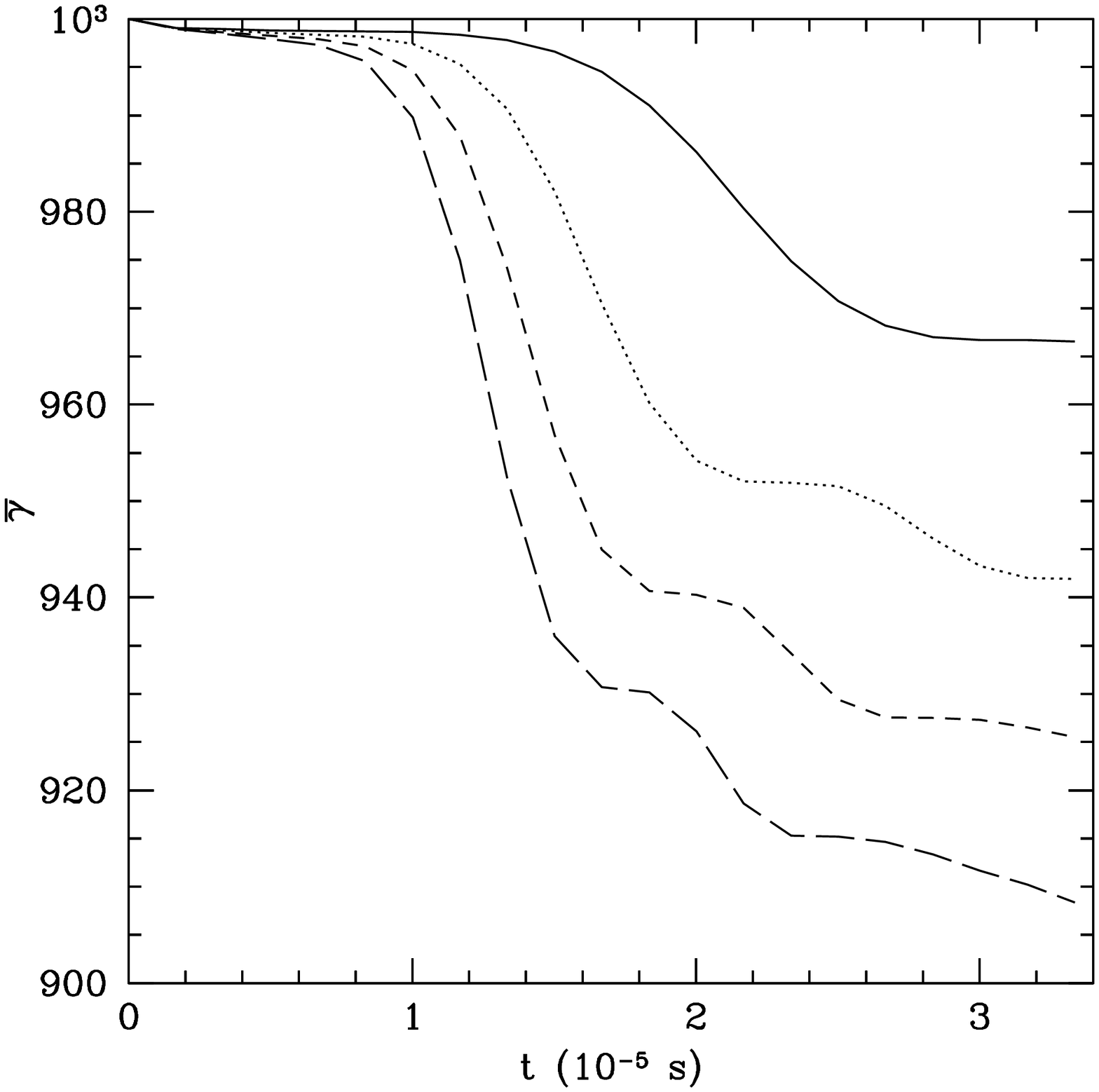}
    \mycaption c{Run 6: From top to bottom: Average Lorentz factor per
    particle, for number density of the bunch $n = 0.1$, 0.2,
    0.3, $0.4\:n_{\textrm{GJ}\ast}$ (labelling as in Fig.~{\ref{fig:n200nfac4_spec}}). Clearly, a small beam number density results in negligible losses, and therefore, no coherent emission. }
    \label{fig:n200nfac4_tgam}
    \end{figure}


\section{Discussion}
We have investigated a specific form of a single-pass free-electron laser process as a possible mechanism
to produce high brightness radio emission of pulsars. We have investigated
the operation of a FEL in the presence of a wiggler which consists of a
  transverse electromagnetic disturbance as it is being overtaken by a
relativistic electron beam.
\\[\baselineskip]
We have shown that, in principle, a FEL can operate in the pulsar outer magnetosphere in the presence of a transverse wiggler. The deciding factors for particles to bunch and emission to be coherent within $3.3 \cdot 10^{-5}\:\mbox{s}$, are the following: a large beam particle density $n \geq 0.1\:n_{\textrm{GJ}}(r_\ast)$, a Lorentz factor of the beam particles $\gamma > 100$ for $K = 10$, and a small background magnetic field $B_0 \leq 10^{-2}\:\mbox{T}$. The brightness temperature of the pulse depends sensitively on these parameters (Table~\ref{tabel:inout}). \\
The required beam particle density together with the wiggler's parameters mainly determine the timescale on which particles start to bunch. Bunching occurs after about 10 times the transit time through the wiggler (i.e. 10 $t_{\textrm{eff}}$,
eq.~\ref{eq:teff}). This seem to set the level of inchorent radiation, which is then large enough for the ponderomotive force to act on the beam particles. 
Then the timescale over which a bunch stays together is about one $t_{\textrm{eff}}$, i.e. coherent radiation only occurs during this period. Due to an increase in axial velocity spread, debunching occurs. Although particles
{\it rebunch}, the associated pulse is weaker due to the velocity
spread in the bunch. In the pulse profile, the first bunching results in one pulse. Then for larger times, particles bunch more often and the pulse acquires more peaks, which are less powerfull.  \\ 
The overall fractional energy losses of the beam are $\sim 5\%$. Together
with the travel time, the calculations demonstrate that when applied to the pulsar magnetosphere for $B \leq 10^{-2}\:\mbox{T}$, the FEL interaction can produce coherent radiation. For a dipole model, this implies $R \gg 2 \cdot 10^3 R_\ast$, where we used $B_\ast = 10^8 \:\mbox{T}$, i.e. in the outer magnetosphere. \\
While the particles are bunched, most radiation is at $\nu_{\textrm{res}}$,
which in our runs varies from $0.7\:\mbox{GHz}$ to $90\:\mbox{GHz}$ as in the pulsar radio emission. 
The lower frequency is obtained with $\gamma = 250$. \\
The particles lose their energy most efficiently for $K=1$. \\
The measured FWHM bandwidth for the radiation $\Delta\nu$ ranges from $0.53\:\mbox{GHz}$ to $14.5\:\mbox{GHz}$, and is broadband $\Delta \nu/\nu \simeq 0.07 - 0.5$. The width of the spectrum highly depends on the number
of bunching events occurring during the simulation. After the first bunching,
the bandwidth is of the order of $1.5\:\mbox{GHz}$; After the second -less
effective- bunching, $\Delta \nu \approx 2.9\:\mbox{GHz}$ and so
on. Of course, the central radiation frequency agrees with radio
pulsar emission by construction. However, it is interesting that the
bandwidth is found to agree
with pulsar radio observations where average pulses and microstructures 
are observed from about $100\:\mbox{MHz}$ to more than $10\:\mbox{GHz}$ (average pulses: e.g. \cite{lm1988,mlabjlhnb1996,lmlbatjbn1998,asbmbnm1998}, microstructures: e.g. \cite{rhc1975,bs1978}). \\
As for the brightness temperature $T_{\textrm{b}}$, which is a big
obstacle for most radiation processes, we found a brightness
temperature $T_{\textrm{b}}$ at pulse maximum of $10^{30}\:\mbox{K}$ for $\beta_{\textrm{w}} = 0.9$
and $T_{\textrm{b}} = 10^{29}\:\mbox{K}$ for a wiggler phase velocity $\beta_{\textrm{w}} \lesssim
0.8$, similar to the observationally derived pulsar brightness
temperatures. \\
Apparently, the FEL is able to
produce the required high brightness pulse. Further, the characteristic opening angle, given by $\tan \phi \approx \lambda/(2 \pi w_0)$, is
$\phi \lesssim 1^\circ$. Again, this agrees with the observed values in pulsar radio
emission, as estimated from the microstructure duration relative to the
pulsar rotation period: 
$\sim 100 \mu \mbox{s}/P \times 360^\circ 
\lesssim 1^\circ$ for $P = 0.1 \:\mbox{s}$ . \\
The radiation pulse has a duration of $2\:\mbox{ns}$. The
shortest elements of radio emission measured from pulsars are 
microstructures. These are quasi-periodic structures of $\sim 10^2
\mu\mbox{s}$, which are broadband and highly (linearly)
polarized \citep{rhc1975,ch1979,lkwj1998,pbcnka2002}. \\
Since the starting time for the particles to bunch is $10\: t_{\textrm{eff}} =
10\lambda_{\textrm{w}}/(\beta_z - \beta_{\textrm{w}})$, the faster the wiggler wave is, the longer it
takes for the particles to get bunched. Also, the characteristic pulse duration $t_{\textrm{eff}}$
increases. This is clearly shown in Run~4, where we
varied the phase velocity of the electromagnetic disturbance between
$0.2c$ and $0.9c$. The computation time limitation forced us to
consider only such relatively low values for electromagnetic wigglers 
in the relativistic outflow from a pulsar magnetosphere, but one can see
from the particles' behaviour in Figure~\ref{fig:N200vfac_tgam},
that even more relativistic electromagnetic disturbances would lead to
longer timescales, e.g. for $\gamma_{\textrm{w}} = (1 - \beta_{\textrm{w}}^2)^{-1/2}$ = 100,
the timescale would go up by a factor 5000, and become comparable to
the observed micropulse durations. (The same argument holds for a lower beam number density $n$, Fig.~{\ref{fig:n200nfac4_tgam}}; i.e. when $n < 0.1 n_{\textrm{GJ}\ast}$, bunching occur at $t > 3.3 \cdot 10^{-5}\:\mbox{s}$). 
\\[\baselineskip]
Finally, we note that the coherent emission from electrons and positrons in a {\it transverse} wiggler add constructively. Therefore, the total number of electrons and positrons in a bunch, and not the charge excess, determines the emission.   
\\[\baselineskip]
In summary, the operation of a single-pass high-gain FEL with a transverse electromagnetic wiggler within the pulsar magnetosphere in the radio regime, requires a mono-energetic beam of electrons/positrons at moderate Lorentz factors in a sufficiently small background magnetic field. 
\\[\baselineskip]
We expect a FEL process to be also possible much nearer to the
pulsar, where the magnetic field strength is
large and the dynamics of the particles are
one-dimensional, when {\it longitudinal} instead of transverse wigglers are used. This would be
the domain of single-pass, high-gain Cerenkov FELs, and a next logical
step to study. Strong Langmuir turbulence has been studied in this context by
\citep{srkl2002}. By using a Particle-in-Cell method, these authors show that electron
scattering on Langmuir turbulence, which is excited in a self-consistent
way, results in high power output of radiation.  
\\[\baselineskip]
{\bf Acknowledgement}: This work has been funded under a joint research
project between the Centre for Plasma Physics and Radiation Technology
(CPS) and the Netherlands Research School for Astronomy (NOVA). We want to thank the anonymous referee for his/her comments. JK likes to thank Dr. Marnix van der Wiel and Dr. Theo Schep for stimulating interest in the project. P.-K. F likes to thank Bas van de Geer, Marieke de Loos for their support in using the code GPT and Cyrille Thomas for the description of the Gaussian modes. Also, we want to thank Kees van der Geer for setting up the simulations. 
\appendix
\section{Derivation of Gaussian modes}
The full derivation of the Gaussian modes can be found in \citet{lasers} and \citet{es1994}. As the radiation is produced by a narrow beam of relativistic particles the radiation propagates mainly in {\it one} direction ($z$-axis). 
As to the dependence of the radiation in the ($x$,$y$)-plane, we assume that, similar to a laser beam, the intensity has a Gaussian form inside the FEL; i.e. $I_\perp \propto e^{-(r_\perp/w_0)^2}$ where $r_\perp = \sqrt{x^2 + y^2}$ and $w_0$ denotes the transverse distance where the intensity drops to $1/e$ of the peak value at $z=0$ ($w_0$ is called the {\it waist}). \\  
The electromagnetic field which has these properties and satisfies Maxwell's equations is the {\it Gaussian mode}. 
The derivation of Gaussian modes starts from Maxwell's
equations in vacuum, which 
result in the (vector) wave equation: 
    \begin{equation}
    \nabla^2 {\bf E}({\bf r}, t) - \frac{1}{c^2}\frac{\partial^2 {\bf
    E}({\bf r}, t)}{\partial t^2} = 0 
    \end{equation}
For vacuum waves ($\omega^2 = k^2c^2$) propagating mainly along
the $z$-direction, the {\it paraxial approximations} can be applied: 
    \begin{eqnarray}
    \left| \frac{\partial^2 {\bf E}({\bf r})}{\partial z^2}\right| &
    \ll & 2 k \left| \frac{\partial {\bf E}({\bf r})}{\partial
    z}\right| \\ 
    \left| \frac{\partial^2 {\bf E}({\bf r})}{\partial z^2}\right| & \ll & 2k \left| {\bf E}({\bf r})\right| 
    \end{eqnarray}
With these, the vector wave equation becomes the vector paraxial wave equation: 
    \begin{equation}
    \nabla^2_{\textrm{T}} {\bf E}({\bf r}) + 2ik \frac{\partial {\bf E}({\bf
    r})}{\partial z} = 0 
    \label{eq:vecwaveequation}
    \end{equation}
where $\nabla^2_{\textrm{T}} = \partial^2/\partial x^2 + \partial^2/\partial y^2$. 
This equation is valid for each component of ${\bf E}({\bf r}) =
E_x({\bf r})\hat{x} + E_y({\bf r})\hat{y} + E_z({\bf r})\hat{z}$. The
same holds for ${\bf B}({\bf r})$. \\
The general solution for each component of this equation, which is
axisymmetric, is the following: 
    \begin{equation}
    u(r_\perp, z) = a \exp(-P(z)) \exp(-\frac{ik r_\perp^2}{2 Q(z)}) 
    \label{eq:gensolution}
    \end{equation}
where $a$ is an amplitude and $P(z)$ and $Q(z)$ are complex functions
which specify the longitudinal and transverse mode behaviour. These can
be retrieved by putting (\ref{eq:gensolution}) into
(\ref{eq:vecwaveequation}), which results in two differential
equations for $P(z)$ and $Q(z)$:  
    \begin{eqnarray}
    \frac{\mbox{d}Q}{\mbox{d}z} & = & 1 \\
    \frac{\mbox{d}P}{\mbox{d}z} & = &\frac{i}{Q}
    \end{eqnarray}
The solutions are simply given by: 
    \begin{eqnarray}
    Q(z) & = & z + q_0 \\
    P(z) & = & -i \ln (z + q_0) 
    \end{eqnarray}
where $q_0$ is an integration constant. Recall that function $Q(z)$
 gives the transverse behaviour of the mode and that the solution is
 paraxial. Therefore, $Q(z)$ can be written in
 terms of a radius of curvature $R(z)$ and a width $w(z)$: 
    \begin{equation}
    \frac{1}{Q(z)} = \frac{1}{z + q_0} = \frac{1}{R(z)} + \frac{-2i}{k w^2(z)}
    \end{equation} 
Furthermore, we assume that at a reference point $z=0$, the mode
wavefront curvature is $R(0) \equiv \infty$, so that: 
    \begin{equation}
    q_0 = \frac{ikw^2(0)}{2}
    \end{equation}
Together with: 
    \begin{eqnarray}
    P(z) & = &-i\ln(z + i z_0) = -i\ln\left[(z^2 +
      z_0^2)e^{i\phi}\right] \\
         & = & -i\ln[z^2 + z_0^2] + \arctan(z_0/z)\\
    z_0 & = & \frac{k w_0^2}{2} \\
    \tan\phi & = & \frac{z_0}{z}
    \end{eqnarray}
the general solution becomes: 
    \begin{equation}
    \begin{split}
    u(r_\perp, z) &= A \frac{w_0}{w(z)} \times \\
& \mathrm{Re}\left[\mathrm{e}^{\frac{-r_\perp^2}{w^2(z)} - i kz +
i \omega t-
    i k \frac{r_\perp^2}{2 R(z)} - i \phi}\right]
    \end{split}
    \end{equation}
where 
    \begin{eqnarray}
    w(z) & = & w_0 \sqrt{1 + \left(\frac{\lambda z}{\pi w_0^2}\right)^2}
\\
    R(z) & = & z + \frac{1}{z}\left( \frac{\pi w_0^2}{\lambda}\right)^2\\
    \tan\phi & = & \frac{\lambda z}{\pi w_0^2} 
    \end{eqnarray}

where $r_\perp$ and $w_0$ are as defined above, $w(z)$ is the waist at $z$ and $A, k, \omega$ are the amplitude, the $z$-component of the wave vector and the frequency of the electromagnetic wave respectively. 
The curvature radius of the wavefront is given by $R(z) = z +
z_0^2/z$, where $z_0 = (\pi w_0^2)/\lambda$ is (roughly) the
separation between the near and the far field. The wavefronts of this mode
change from planar in the near field to spherical in the far field
(see Fig.~\ref{fig:nearfar}). 
\\[\baselineskip]
The opening angle $\psi$ of the {\it intensity} of the mode is given by: 
    \begin{equation}
    \tan{\phi} = \frac{1}{2}\sqrt{2}\frac{w(z)}{z} = \sqrt{\frac{\lambda}{2 \pi z_0}}
    \label{eq:openangle1}
    \end{equation}
Within distance of length $z_0$, wavefronts of the Gaussian modes
are considered planar. Different wavelengths and different $z_0$
result in different opening angles. In table~\ref{tabel:openhoek} is
listed the range of opening angles when we consider radio waves with
wavelengths between 3~cm to 30~m and some typical distances in the pulsar magnetosphere.  
    \begin{table}[h!]
    \centering
    \begin{tabular}{|ll|}
    \hline
    $z_0$ &  $\phi$ \\
    \hline
    polar cap radius $R_{\textrm{pc}}$ = 100 m & 24' - 3$^\circ$57'\\
    stellar radius $R_{\ast}$ = $10^4$ m & 2'23'' - 24'\\
    light cilinder $R_{\textrm{lc}}$ = $10^7$ m & 4''.5 - 45''\\ 
    \hline
    \end{tabular}
    \caption{The range of opening angles $\phi$ for different $z_0$
    and wavelengths between 3 cm and 30 m according to
    equation~(\ref{eq:openangle1}). }
    \label{tabel:openhoek}
    \end{table}


\bibliography{0295ref}  
\bibliographystyle{aa}
\end{document}